\PassOptionsToPackage{dvipsnames}{xcolor}
\documentclass[acmsmall,screen,nonacm]{acmart}
\pdfoutput=1
\setcopyright{none}

\usepackage{booktabs}   %
\usepackage{subcaption} %

\usepackage[export]{adjustbox}
\usepackage[normalem]{ulem}
\usepackage[scaled=0.8]{beramono}
\usepackage{placeins}
\usepackage[skip=2.5pt]{caption}
\usepackage{bcprules}
\usepackage{bm}
\usepackage{cancel}
\usepackage{diagbox}
\usepackage{enumitem}
\usepackage{etoolbox}
\usepackage{float}
\usepackage{lipsum}
\usepackage{listings,array,varwidth}
\usepackage{makecell}
\usepackage{mathpartir}
\usepackage{amsmath}
\usepackage{mathtools}
\usepackage{mdframed}
\usepackage{multirow}
\usepackage{multifootnote}
\usepackage{needspace}
\usepackage{scalerel}
\usepackage{slashbox}
\usepackage{soul}
\usepackage{tabularx}
\usepackage{tikz-cd}
\usepackage{tikz} %
\usepackage{varioref}
\usepackage{wrapfig}
\usepackage{xcolor,colortbl}
\usepackage{xfrac}
\usepackage{xifthen}
\usepackage{xspace}
\usepackage{xurl}
\usetikzlibrary{fit, calc, matrix, arrows, arrows.meta, shapes, positioning}
\tikzset{
  cellBase/.style = {draw, minimum size=4.5mm, inner sep=0pt, font=\scriptsize, align=center, text width=4.5mm,anchor=center,
                     fill=gray!5, draw=gray!60},
  cell/.style     = {cellBase, fill=#1!18, draw=#1!60},
  cell/.default   = gray,
  label/.style = {anchor=east, font=\scriptsize, text=#1},
  label/.default = black,
  codefont/.style={font=\fontsize{11pt}{13pt}\selectfont\ttfamily},
  code/.style     ={codefont, text=#1, align=center, anchor=center},
  code/.default   = blue,
  annote/.style   = {font=\scriptsize, text=#1},
  annote/.default = black,
  outarrow/.style = {-Stealth, semithick, draw=#1},
  outarrow/.default = black,
  interarrow/.style = {-Stealth, semithick, draw=#1},
  interarrow/.default = black,
  swaparrow/.style = {<->, Stealth-Stealth, semithick, draw=#1},
  swaparrow/.default = black,
  midcircle/.style={midway, circle, draw, fill=#1, minimum size=3pt, inner sep=0pt},
  middiamond/.style={midway, diamond, aspect=1, draw, fill=#1, inner sep=1pt, minimum size=3pt},
  midlabel/.style={font=\scriptsize, color=#1, inner sep=.2ex}, %
  title/.style    = {font=\scriptsize, anchor=south},
}

\usepackage{cleveref}
\usepackage{pifont} %

\colorlet{outerovalcolor}{blue!5}
\colorlet{innerovalcolor}{blue!15}

\setlist[enumerate, 1]{%
  leftmargin = 1.2\parindent, %
  align = left,
  labelwidth=\parindent,
  labelsep = 1pt
}
\lstdefinelanguage{DOT}%
{morekeywords={val,new},%
  sensitive,%
  morecomment=[l]//,%
  morecomment=[s]{/*}{*/},%
  morestring=[b]",%
  morestring=[b]',%
  showstringspaces=false%
}[keywords,comments,strings]%

\newlength{\trulemargin}
\newlength{\trulewidth}
\newlength{\srulewidth}
\newenvironment{trules}{$\vspace{0.5em}\ba{p{\trulemargin}@{~}p{\trulewidth}@{~}p{\trulemargin}}}{\ea$}
\newenvironment{srules}{$\vspace{0.5em}\ba{p{\trulemargin}@{~}p{\srulewidth}}}{\ea$}

\newcommand{\ba}{\begin{array}}
\newcommand{\ea}{\end{array}}

\newcommand{\ei}{\end{array}}
\newcommand{\bcases}{\left\{\begin{array}{ll}}
\newcommand{\ecases}{\end{array}\right.}

\newcommand{\eg}{{\em e.g.}\xspace}
\newcommand{\ie}{{\em i.e.}\xspace}

\newcommand{\dom}{\mbox{\sl dom}}

\newcommand{\judgement}[2]{{\textsf{\textbf{#1}}} \hfill #2}

\usepackage{listings,array,varwidth}
\usepackage{mathpartir}
\usepackage{mathtools}
\usepackage{makecell}

\usepackage{diagbox}
\usepackage{float}
\usepackage{varioref}
\usepackage{xfrac}
\usepackage{enumitem}
\usepackage[skip=2.5pt]{caption}
\usepackage{slashbox}

\usepackage{xifthen}

\usepackage{tikz}
\usepackage{xparse}
\usepackage{xsavebox}
\usetikzlibrary{arrows}

\makeatletter %
\def\arcr{\@arraycr}
\makeatother

\newcommand{\showDOI}[1]{\unskip}

\newtheorem{innercustomgeneric}{\customgenericname}
\providecommand{\customgenericname}{}
\newcommand{\newcustomtheorem}[2]{%
  \newenvironment{#1}[1]
  {%
   \renewcommand\customgenericname{#2}%
   \renewcommand\theinnercustomgeneric{##1}%
   \innercustomgeneric
  }
  {\endinnercustomgeneric}
}
\newcustomtheorem{re-definition}{Definition}
\newcustomtheorem{re-lemma}{Lemma}

\makeatletter
\DeclareRobustCommand{\etc}{%
    \@ifnextchar{.}%
        {etc}%
        {etc.\@\xspace}%
}
\makeatother

\newcommand{\Specsharp}{%
	{\settoheight{\dimen0}{C}Spec\kern-.05em \resizebox{!}{\dimen0}{\raisebox{\depth}{\#}}}}
\newcommand{\Csharp}{%
	{\settoheight{\dimen0}{C}C\kern-.05em \resizebox{!}{\dimen0}{\raisebox{\depth}{\#}}}}

\newcommand{\fun}[1]{\operatorname{#1}}
\newcommand{\DOM}{\fun{dom}}

\definecolor{blue-violet}{rgb}{0.54, 0.17, 0.89}
\definecolor{dark-cyan}{HTML}{135579}
\definecolor{magenta}{HTML}{a8264f}

\colorlet{mute}{teal}
\colorlet{eff}{magenta}
\newcommand{\mute}[1]{{\color{mute}#1}}

\definecolor{light-gray}{gray}{0.92}
\definecolor{dark-gray}{gray}{0.5}
\definecolor{light-yellow}{HTML}{FFFACD}

\newcommand{\commentstyle}{\color{dark-cyan}}
\newcommand{\errstyle}{\color{red}}
\newcommand{\lstcm}[1]{{\commentstyle#1}} %
\newcommand{\lsterrcm}[1]{{\errstyle#1}}

\lstdefinelanguage{LambdaCirc}%
{morekeywords={abstract,%
  case,catch,char,class,%
  def,else,extends,final,finally,for,%
  if,import,implicit,%
  match,module,%
  new,null,%
  object,override,%
  package,private,protected,public,%
  for,public,return,super,%
  this,throw,trait,try,type,%
  val,var,%
  with,while,%
  yield,%
  let,end,%
	in,fun,alloc,inc,%
  nil,cons%
  },%
  mathescape=true,%
  sensitive,%
  keywordstyle={\color{dark-cyan}\bf\ttfamily},%
  commentstyle=\commentstyle,%
  morecomment=[l]//,%
  morecomment=[s]{/*}{*/},%
  morecomment=[s][\color{dark-cyan}]{@}{\ },%
  morestring=[b]",%
  morestring=[b]',%
  showstringspaces=false%
}[keywords,comments,strings]%

\newcommand{\trackset}[1]{^{\texttt{\{#1\}}}}

\newcommand{\trackvar}[1]{^{\texttt{#1}}}

\newcommand{\maybelang}{\ensuremath{\mathsf{F}_{\varepsilon <:}^{\vardiamondsuit}}\xspace}

\makeatletter
\def\ifenv#1{%
   \def\@tempa{#1}%
   \ifx\@tempa\@currenvir
      \expandafter\@firstoftwo
    \else
      \expandafter\@secondoftwo
   \fi
}
\makeatother

\makeatletter
\edef\showenv{\@currenvir}
\makeatother

\newcommand{\Type}[1]{\ensuremath{\ifenv{lstlisting}{\texttt{#1}}{\mathsf{#1}}}}
\newcommand{\typevar}[1]{\ensuremath{\ifenv{lstlisting}{\texttt{#1}}{#1}}}
\newcommand{\ty}[2][]{\ensuremath{\ifthenelse{\isempty{#1}}{\typevar{#2}}{\typevar{#2}^{\,\typevar{#1}}}}}

\newcommand{\Var}{\Type{Var}}

\newcommand{\TRef}{\Type{Ref}}
\newcommand{\TRefpow}[2]{\TRef^{#1}\ifenv{lstlisting}{\,#2}{\,#2}}
\newcommand{\TTop}{\top}
\newcommand{\TUnit}{\Type{Unit}}
\newcommand{\Loc}{\Type{Loc}}

\newcommand{\TBool}{\Type{Bool}}
\newcommand{\TList}{\Type{List}}

\newcommand{\tunit}{\text{\textbf{\textsf{unit}}}}
\newcommand{\turef}{\text{\textbf{\textsf{uref}}}}
\newcommand{\tref}{\text{\textbf{\textsf{ref}}}}
\newcommand{\tswap}{\text{\textbf{\textsf{swap}}}}
\newcommand{\tget}{\text{\textbf{\textsf{get}}}}

\newcommand{\tnil}{\text{\textbf{\textsf{nil}}}}
\newcommand{\tcons}{\text{\textbf{\textsf{cons}}}}

\newcommand{\tput}{\text{\textbf{\textsf{put}}}}
\NewDocumentCommand{\tifnil}{ G{\ensuremath{\cdots}} G{\ensuremath{\cdots}} G{\ensuremath{\cdots}} }{\text{\textbf{\textsf{ifnil}}}~#1~\text{\textbf{\textsf{then}}}~#2~\text{\textbf{\textsf{else}}}~#3}
\NewDocumentCommand{\tmatchlist}{ g G{\ensuremath{\cdots}} G{\ensuremath{\cdots}} }{%
  \IfNoValueTF{#1}%
    {\text{\textbf{\textsf{match}}}~\ensuremath{\cdots}~\text{\textbf{\textsf{with}}}~\ensuremath{\cdots}}%
    {\text{\textbf{\textsf{match}}}~#1~\text{\textbf{\textsf{with}}}~\tnil\to #2 ;~ \tcons~#3}%
}

\newcommand{\tfree}{\text{\textbf{\textsf{free}}}}
\newcommand{\tmove}{\text{\textbf{\textsf{move}}}}
\NewDocumentCommand{\trefat}{ G{\ensuremath{\cdots}} G{\ensuremath{\cdots}} }{\ensuremath{\mathsf{ref}~#1~\mathsf{at}~#2}}

\NewDocumentCommand{\tsucc}{ G{\ensuremath{\cdots}} }{\Type{succ}~#1}
\NewDocumentCommand{\tpred}{ G{\ensuremath{\cdots}} }{\Type{pred}~#1}
\NewDocumentCommand{\tmul}{ G{\ensuremath{\cdots}} G{\ensuremath{\cdots}} }{\Type{mul}~#1~#2}
\NewDocumentCommand{\tiszero}{ G{\ensuremath{\cdots}} }{\Type{iszero}~#1}
\NewDocumentCommand{\tif}{ G{\ensuremath{\cdots}} G{\ensuremath{\cdots}} G{\ensuremath{\cdots}} }{\Type{if}~#1~\Type{then}~#2~\Type{else}~#3}

\NewDocumentCommand{\ext}{s m}{\IfBooleanTF{#1}{\HLBox[blue!10]{#2}}{\HLBox[gray!20]{#2}}}

\newcommand{\SR}[2]{\TRef\ifenv{lstlisting}{}{~}[\ty[#2]{#1}]}

\newcommand{\mty}[3][x]{\ensuremath{\ty[#2]{#3}}}

\newcommand{\mr}[4][x]{\mty[#1]{#2}{\SR{#3}{#4}}}

\newcommand{\TAll}[7]{\ensuremath{\forall f(\ty[#2]{#1} <: \ty[#4]{#3}) .^{#7} \ty[#6]{#5}}}
\newcommand{\TLam}[5]{\ensuremath{\Lambda f(\ty[#2]{#1}) . {#5}}}
\newcommand{\TApp}[3]{\ensuremath{{#1}\ [\ty[#3]{#2}]}}

\newcommand{\ts}[1][]{\ensuremath{\ifthenelse{\isempty{#1}}{\,\vdash\,}{\,\vdash^{\,#1}\,}}}

\newcommand{\flt}{\ensuremath{\varphi}}

\newcommand{\cx}[2][]{\ensuremath{\ifthenelse{\isempty{#1}}{#2}{#2^{\,#1}}}}

\newcommand{\csx}[3][]{\ensuremath{\ifthenelse{\isempty{#1}}{#3\mid #2}{{\color{gray!50}[}#3\mid#2{\color{gray!50}]}^{\,#1}}}}

\providecommand{\G}{G} %
\renewcommand{\G}[1][]{\cx[#1]{\Gamma}}

\newcommand{\HLBox}[2][teal!12]{\ensuremath{\mathchoice%
  {\setlength{\fboxsep}{.5ex}\colorbox{#1}{$\displaystyle#2$}}%
  {\setlength{\fboxsep}{.5ex}\colorbox{#1}{$\textstyle#2$}}%
  {\setlength{\fboxsep}{.5ex}\colorbox{#1}{$\scriptstyle#2$}}%
  {\setlength{\fboxsep}{.5ex}\colorbox{#1}{$\scriptscriptstyle#2$}}}}%

\newcommand{\QFresh}{\ensuremath{\vardiamondsuit}}
\newcommand{\qbot}{\ensuremath{\varnothing}}
\newcommand{\qfresh}{\ensuremath{\vardiamondsuit}}
\newcommand{\subq}{\ensuremath{\subseteq}}
\newcommand{\qlub}{\ensuremath{\cup}}
\newcommand{\qglb}{\ensuremath{\cap}}

\xsavebox*{SMALLSTAR}{\(\raisebox{.25ex}{\(\qfresh\)}\)}

\xsavebox*{OVRLP}{$\raisebox{.37ex}[0pt][0pt]{$\mathrlap{\hspace{.415ex}\scaleobj{.5}{\vardiamondsuit}}$}\cap$}
\def\overlap{\ensuremath{\mathbin{\scalerel*{\xusebox{OVRLP}}{\sqcap}}}}

\newcommand{\WF}[1]{\ensuremath{#1\ \mathsf{ok}}}
\newcommand{\reaches}{\ensuremath{\mathrel{\leadsto}}}
\newcommand{\norm}[1]{\lvert #1 \rvert}

\newcommand{\qtrans}[2][]{\ensuremath{\ifthenelse{\isempty{#1}}{#2\mathord{*}}{{#2}^{#1}}}}

\newcommand{\BOX}[1]{\fbox{$\strut #1$}}

\newcommand{\FV}{\ensuremath{\operatorname{fv}}}
\newcommand{\FTV}{\ensuremath{\operatorname{ftv}}}

\newcounter{typerule}
\crefformat{typerule}{#2\textsc{#1}#3}
\Crefformat{typerule}{#2\textsc{#1}#3}

\newcommand{\typerule}[3]{%
  \def\thetyperule{#1}%
  \refstepcounter{typerule}%
  \label{typing:#1}%
  \infrule[#1]{#2}{#3}
}
\newcommand{\vgap}{\vspace{7pt}}

\newcommand{\hole}[1]{\ensuremath{[\,#1\,]}}
\newcommand{\CX}[3][black]{\ensuremath{{\color{#1}#2\ifthenelse{\isempty{#3}}{}{\hole{{\color{black}#3}}}}}}

\newcounter{semrule}
\crefformat{semrule}{#2\textsc{#1}#3}
\Crefformat{semrule}{#2\textsc{#1}#3}

\makeatletter

\newcommand{\rulename}{\@ifstar\rulename@star\rulename@nostar}
\newcommand{\rulename@nostar}[2][]{%
  \def\rulename@id{#2}%
  \ifthenelse{\isempty{#1}}{}{%
    \def\rulename@id{#1}%
  }%
  \@ifundefined{rulename@\rulename@id}{%
    \expandafter\gdef\csname rulename@\rulename@id\endcsname{}%
    \def\thesemrule{\rulename@id}%
    \refstepcounter{semrule}%
    \label{rule:\rulename@id}%
  }{%
    \def\thesemrule{\rulename@id}%
  }%
  (\textsc{#2})%
}
\newcommand{\rulename@star}[1]{(\textsc{#1})}
\makeatother

\newcommand{\semrulelabel}[1]{%
  \begingroup
    \def\thesemrule{#1}%
    \refstepcounter{semrule}%
    \label{rule:#1}%
  \endgroup
}

\newcommand{\FX}[1]{\ensuremath{{\color{eff}#1}}}
\newcommand{\EPS}[1][]{\ifthenelse{\isempty{#1}}{\FX{\varepsilon}}{\FX{\varepsilon_{#1}}}}
\newcommand{\EPSPR}[1][]{\ifthenelse{\isempty{#1}}{\FX{\varepsilon'}}{\FX{\varepsilon'_{#1}}}}
\newcommand{\PURE}{\FX{\varnothing}}
\newcommand{\NOTEFFSEQ}{\ensuremath{\mathbin{\FX{\ntriangleright}}}}
\newcommand{\EFFSEQ}{\ensuremath{\mathbin{\FX{\rhd}}}}
\newcommand{\EFFJOIN}{\ensuremath{\mathbin{\FX{\sqcup}}}}

\newcommand{\USECOMP}[1]{\ensuremath{\mathscr{u}(#1)}}
\newcommand{\KILLCOMP}[1]{\ensuremath{\mathscr{k}(#1)}}

\newcommand{\USE}{\ensuremath{\mathscr{u}}}
\newcommand{\KILL}{\ensuremath{\mathscr{k}}}

\newcommand{\UKX}[2]{\FX{\{\USE: #1;\, \KILL: #2\}}}

\newcommand{\JUSTU}[1]{\FX{\{\USE: #1\}}}
\newcommand{\JUSTK}[1]{\FX{\{\KILL: #1\}}}

\newcommand{\JUSTM}[1]{\UKX{#1}{#1}}

\newcommand{\DEP}[1][]{\ensuremath{\ifthenelse{\isempty{#1}}{\mute{\delta}}{\mute{\delta_{#1}}}}}

\newcommand{\HDEP}[1][]{\ensuremath{\ifthenelse{\isempty{#1}}{\mute{\mathsf{h}}}{\mute{\mathsf{h}_{#1}}}}}
\newcommand{\SDEP}[1][]{\ensuremath{\ifthenelse{\isempty{#1}}{\mute{\mathsf{s}}}{\mute{\mathsf{s}_{#1}}}}}

\newcommand{\redv}{\ensuremath{\mathrel{\longrightarrow}}}

\newcommand{\DEAD}{\text{\ding{61}}}

\graphicspath{ {./images/} }
\usepackage[normalem]{ulem}
\makeatletter
\UL@protected\def\reduwave{\leavevmode \bgroup
\ifdim \ULdepth=\maxdimen \ULdepth 3.5\p@
\else \advance\ULdepth2\p@
\fi \markoverwith{\lower\ULdepth\hbox{\textcolor{red}{\sixly \char58}}}\ULon}
\makeatother

\usepackage{xr}
 
\makeatletter%
\begin{document}

\title{Capture Now, Consume Later: Reachability Types with Flow-Sensitive Effects for Higher-Order Ownership Transfer}

\author{Haotian Deng}
\orcid{0009-0002-7096-2646}
\affiliation{%
  \institution{Purdue University}
  \city{West Lafayette}
  \country{USA}
}
\email{deng254@purdue.edu}

\author{Siyuan He}
\orcid{0009-0002-7130-5592}
\affiliation{%
  \institution{Purdue University}
  \city{West Lafayette}
  \country{USA}
}
\email{he662@purdue.edu}

\author{Songlin Jia}
\orcid{0009-0008-2526-0438}
\affiliation{%
  \institution{Purdue University}
  \city{West Lafayette}
  \country{USA}
}
\email{jia137@purdue.edu}

\author{Yuyan Bao}
\orcid{0000-0002-3832-3134}
\affiliation{%
  \institution{Augusta University}
  \city{Augusta}
  \country{USA}
}
\email{yubao@augusta.edu}

\author{Tiark Rompf}
\orcid{0000-0002-2068-3238}
\affiliation{%
  \institution{Purdue University}
  \city{West Lafayette}
  \country{USA}
}
\email{tiark@purdue.edu}

\lstMakeShortInline[keywordstyle=,%
                    flexiblecolumns=false,%
                    language=Scala,
                    basewidth={0.56em, 0.52em},%
                    mathescape=false,%
                    basicstyle=\footnotesize\ttfamily]@

\begin{abstract}
Higher-order impure programs routinely transfer or consume a resource
\emph{through a closure}: when one access path consumes the resource, it must be
disabled through all remaining aliases---including those captured in
closures---precisely at the point of consumption. We present a flow-sensitive
type-and-effect system for reachability types that makes such ownership transfer
sound in the presence of higher-order control flow. The key idea is simple:
every computation can induce a \textit{use} and/or a \textit{kill} effect. The
system statically tracks them, rejects any use of a killed resource. With
reachability qualifiers and flow-sensitive effects, the system also
\textit{derives} on-demand uniqueness patterns compatible with higher-order
impure functions.

We illustrate the practicality of the system through numerous case studies
such as a freezable counter, a destructive merge on unique linked lists,
Rust-style swap on \texttt{Box}, recursive sum on unique linked lists, and
use-once continuations. We formalize the system as the
\ensuremath{\mathsf{F}_{\varepsilon <:}^{\vardiamondsuit}} calculus, present its
typing rules and operational semantics, and prove effect soundness, including
multi-step preservation and preservation under parallel reduction. We mechanize
all key results in Rocq.

 \end{abstract}


\maketitle

\section{Introduction}
\label{sec:intro}

Closures are a ubiquitous pattern in higher-order functional programming. They
can capture references to shared mutable state and delay execution---often far
from where they were created. Multiple closures may read and write to the same
captured state; eventually, one of them may need to \emph{consume} that state,
transferring its ownership elsewhere or reclaiming it, after which all other
closures must stop using it. The key challenge is statically ensuring the safe
sharing and consumption of resources in the presence of higher-order control
flow.

This challenge spans several distinct lines of research, but none provide the
ideal foundation for higher-order functional programming. Traditional ownership
systems~\citep{DBLP:conf/oopsla/ClarkePN98,DBLP:conf/ecoop/NobleVP98} and
Fearless Concurrency~\citep{DBLP:conf/pldi/MilanoTM22} offer sophisticated alias
control and ownership transfer, and
Cyclone~\citep{Proc.ACMSIGPLAN2002Conf.Program.Lang.Des.Implement./GrossmanGTMY02,DBLP:conf/iwmm/HicksMGJ04}
provides safe manual deallocation, but these approaches all operate strictly in
first-order settings.  Classical region
calculi~\citep{DBLP:conf/popl/TofteT94,DBLP:conf/pldi/AikenFL95,lippmeier_mechanized_2013}
natively support closures but rely on bulk lexical reclamation rather than
flow-sensitive consumption. Meanwhile, capability
systems~\citep{clebsch_deny_2015,Proc.ACMProgram.Lang./ArvidssonESSJM23,ECOOP2010-Object-OrientedProgram./Haller10}
and Rust~\citep{DBLP:journals/pacmpl/0002JKD18} enforce ownership rigidly at
capture time, and substructural
approaches~\citep{DBLP:journals/pacmpl/BernardyBNJS18,ProgrammingLanguagesandSystems/MarshallM22,Proc.ACMProgram.Lang./LorenzenD23}
lack the precision to distinguish between merely mentioning a resource and
actually using it. 

As the widespread adoption of Rust establishes strict ownership model as the
prevailing standard for resource management, a critical gap remains in
reconciling this paradigm with higher-order control flow. First, because 
ownership models forbid shared mutability, closures are often forced to take
strict ownership of captured resources, prematurely transferring ownership at
the moment of capture rather than
execution~(\Cref{sec:intro-ownership-transfer}). Second, ownership models
frequently impose a strict dichotomy that forces developers to choose between
shared mutable aliases and explicit manual deallocation, leaving no safe
mechanism to manually destroy or transfer a resource while it remains
shared~(\Cref{sec:intro-shared-destruction}). Finally, while exclusivity is
essential, built-in uniqueness primitives achieve it only by sacrificing
higher-order control flow or sharing. These rigid invariants scale poorly to
escaping closures, preventing uniqueness from serving as an adaptable, opt-in
mode of use~(\Cref{sec:intro-uniqueness})

Reachability types (RT) offer a unique perspective. Rather than imposing rigid
ownership boundaries, coarse-grained lexical regions, or strict global
linearity, RT \textit{tracks} which variables or locations an expression may
reach. This natively accommodates shared mutable references, side effects, and
higher-order control
flows~\citep{DBLP:journals/pacmpl/BaoWBJHR21,DBLP:journals/pacmpl/WeiBJBR24}.
Recent extensions enrich RT with cyclic references, shallow
tracking~\citep{Proc.ACMProgram.Lang./DengSSY25}, and lexical arena-based
lifetimes~\citep{he_when_2025}. 

Existing RT systems, however, lack a flow-sensitive mechanism to track resource
consumption and disable access. This paper fills that gap with a flow-sensitive
effect system (see \Cref{sec:motivation-effects}) for an impure, higher-order
functional language.  The key idea is simple: every computation can induce a
\textit{use} and/or \textit{kill} effect. The system tracks these effects and
rejects any use of a resource after it has been killed. Function types carry
\textit{latent} effects that are induced upon function application.  We argue
this synthesis provides the ideal foundation for closures, driven by
\textit{three key ingredients} for safe ownership transfer with higher-order
control flow:

\textit{Deferring Consumption.} First, closures must decouple capture from
consumption. Rather than prematurely transferring ownership at creation time, a
closure must disable a shared resource only when executed. Our system achieves
this by carrying consumption as latent effects on function types, and induce the
effect only at call site.

\textit{Disabling Aliases.} Second, shared mutability and manual deallocation
should be permitted to coexist. In other words, developers should not be forced
to choose between shared mutability and deterministic reclamation. By disabling
all tracked aliases upon consumption using flow-sensitive effects, our system
safely supports both.

\textit{Deriving Exclusivity.} Third, uniqueness patterns should be compatible
with higher-order control flow. By deriving exclusivity from reachability
tracking and effects, uniqueness in our system becomes an emergent mode of use
that seamlessly interacts with higher-order control flow.

\smallskip
We now illustrate these three ingredients through a series of motivating examples,
and then summarize the contributions of this paper.

\subsection{Deferring Consumption: Ownership Transfer Under Higher-Order Control Flow}
\label{sec:intro-ownership-transfer}

Consider a freezable counter, as shown in \Cref{fig:intro-counter-freeze}.  The
\texttt{freeze} closure carries a latent \textit{use}/\textit{kill} effect on a
counter's shared state. Invoking it consumes that state, meaning any later call
through a sibling closure that reaches the state (\eg \texttt{incr}) is
statically rejected. This is achieved using the same reachability qualifiers
and \textit{use}/\textit{kill} effects that already track access and
deallocation, now carried as \emph{latent effects on function types} and
induced at application. A \emph{fresh application} yields the fresh remaining
access path.

Because the counter's state is here simply \emph{discarded} rather than
returned, the idempotent \tfree\ operator alone suffices to disable access.
\Cref{sec:motivation-idempotent} introduces its non-idempotent counterpart,
\tmove, which additionally \emph{returns} the payload as a fresh resource while
disabling the old alias, modeling a ownership transfer.

The example rests on a design principle that distinguishes this work: closures
are \emph{not owners} of references. A closure's reachability qualifier simply
records which resources it \emph{may reach}, and its latent effect records what
\emph{calling} it will do to them; the flow-sensitive step---ownership
transfer or deallocation---occurs only when the closure is \emph{called}. Rust,
by contrast, treats a closure primarily as a \textit{data structure} that assumes ownership of the
values it captures, rather than as a functional abstraction of behavior. We
argue that proper functional programming should embrace the latter.

\begin{figure*}[tpb]
  \centering
  \begin{subfigure}[t]{0.42\textwidth}
\begin{lstlisting}[basicstyle=\ttfamily\footnotesize,breaklines=true]
// reusable counter implementation
type Counter = $\mu$p.Pair[
  (Unit =>$^{\JUSTU{\texttt{p}}}$ Unit)$\trackvar{p}$,
  (Unit =>$^{\JUSTM{\texttt{p}}}$ Int)$\trackvar{p}$]
def counter(n: Int): Counter$\trackvar{p}$ = 
  val c = new Ref(n)
  def freeze =  
    val v = !c // : $\JUSTU{\texttt{c}}$
    free(c)    // : Unit$^\lstcm{\qbot}$ $\JUSTK{\texttt{c}}$
    v 
  ({ c := !c + 1 }, freeze)
\end{lstlisting}
    \caption[A freezable counter.]{A freezable counter\footnotemark{}. \texttt{freeze} carries a latent
    \textit{use}/\textit{kill} effect on the shared state \texttt{p}.}
    \label{fig:intro-counter-def}
  \end{subfigure}
  \hfill
  \begin{subfigure}[t]{0.56\textwidth}
\begin{lstlisting}[basicstyle=\ttfamily\footnotesize,breaklines=true]
// user program that uses the counter
val ctr = counter(42) // : Counter$\lstcm{\trackvar{ctr}}$
(incr, frz) = ctr
          // incr : (Unit =>$^{\JUSTU{\texttt{ctr}}}$ Unit)$\lstcm{\trackvar{ctr}}$
          // frz  : (Unit =>$^{\JUSTM{\texttt{ctr}}}$ Int)$\lstcm{\trackvar{ctr}}$
incr()         // : Unit$^\lstcm{\qbot}$ $\JUSTU{\texttt{p}}$             $\lstcm{\leftarrow \text{state} \to 43}$
incr()         // : Unit$^\lstcm{\qbot}$ $\JUSTU{\texttt{p}}$             $\lstcm{\leftarrow \text{state} \to 44}$
val fv = frz() // : Int$^\lstcm{\qbot}$ $\UKX{\texttt{p}}{\texttt{p}}$   $\lstcm{\leftarrow \text{fv = 44; kills p}}$
incr()         // : $\JUSTK{\texttt{p}} \NOTEFFSEQ \JUSTU{\texttt{p}}$ $\lsterrcm{\text{Error: counter killed}}$
frz()          // : $\JUSTK{\texttt{p}} \NOTEFFSEQ \JUSTU{\texttt{p}}$ $\lsterrcm{\text{Error: counter killed}}$
incr()         // : $\JUSTK{\texttt{p}} \NOTEFFSEQ \JUSTU{\texttt{p}}$ $\lsterrcm{\text{Error: counter killed}}$
\end{lstlisting}
    \caption{Using the freezable counter. Invoking \texttt{frz} consumes \texttt{p}; later calls through
    \texttt{frz} or \texttt{incr} are statically rejected.}
    \label{fig:intro-counter-use}
  \end{subfigure}

  \caption{Ownership transfer through a higher-order value.  Calling the
  \texttt{freeze} closure \texttt{frz} induces its latent
  \textit{use}/\textit{kill} effect on the shared state \texttt{p}, so any later
  use of the shared state, even through a \emph{different} closure (\eg
  \texttt{incr}) is statically rejected.}
  \label{fig:intro-counter-freeze}
\end{figure*}
 
\begin{figure}[t]
  \centering
\begin{lstlisting}[basicstyle=\ttfamily\footnotesize,breaklines=true]
let c = Box::new(n);
let freeze = || { let v = *c; drop(c); v }; // c is moved into freeze here, at creation
let incr = || *c += 1; // $\lsterrcm{\text{Error: borrow of moved value \texttt{c}}}$
\end{lstlisting}
  \caption{Rust's failed attempt at the freezable counter of
  \Cref{fig:intro-counter-freeze}. Capture \emph{is} the ownership transfer:
  \texttt{c} moves into \texttt{freeze} when the closure is created, so the
  sibling closure \texttt{incr} is rejected before anything runs. In RT the
  counter typechecks because closures are not owners; the state is consumed
  only when \texttt{freeze} is called.}
  \label{fig:intro-capture-vs-call}
\end{figure}
 
Concretely, in Rust, capture \emph{is} ownership transfer: creating a closure
that consumes a value immediately moves that value into the closure at creation
time. Consequently, as \Cref{fig:intro-capture-vs-call} demonstrates, the
freezable counter cannot be written in safe Rust.  The reference \texttt{c}
moves into \texttt{freeze} upon creation, causing the sibling closure
\texttt{incr} to be rejected by the borrow checker before any execution
occurs. This failure is systemic rather than incidental: deallocating
\texttt{c} requires exclusive ownership (Rust's \texttt{drop} consumes its
operand), yet the counter demands that \texttt{c} be shared among sibling
closures. Because Rust cannot accommodate both simultaneously, consumption is
forced at the point of capture.

\begin{figure*}[tbp]
  \centering
  \begin{subfigure}[t]{0.29\textwidth}
    \begin{lstlisting}[basicstyle=\ttfamily\scriptsize,breaklines=true,language=C++]
llvm::Instruction* p = inst;
llvm::Instruction* q = inst;
auto defer_access = [q]() { 
  llvm::errs() << *q << "\n";
};
p->eraseFromParent();
defer_access();
// $\lsterrcm{\text{Runtime Error: use after free}}$
    \end{lstlisting}
    \caption{Compiler IR APIs may expose aliases where one pointer deletes the shared resource.}
    \label{fig:intro-shared-ir}
  \end{subfigure}
  \hfill
  \begin{subfigure}[t]{0.33\textwidth}
    \begin{lstlisting}[basicstyle=\ttfamily\scriptsize,breaklines=true]
let p = Rc::new(RefCell::new(0));
let q = Rc::clone(&p); // p, q share
let defer_access = move || {
    *q.borrow_mut() += 1; 
};
// drop(p.into_inner());
// p lives until `defer_access` dies.
// $\lsterrcm{\text{can't manually deallocate p}}$
    \end{lstlisting}
    \caption{Rust safely recovers shared mutability via \texttt{Rc}, but strictly forbids manual deallocation.}
    \label{fig:intro-shared-rust}
  \end{subfigure}
  \hfill
  \begin{subfigure}[t]{0.33\textwidth}
    \begin{lstlisting}[basicstyle=\ttfamily\scriptsize,breaklines=true]
val p = new Ref(0) // : Ref[Int]$\lstcm{\trackvar{p}}$
val q = p // : Ref[Int]$\lstcm{\trackvar{q}}$
def defer_access() = {
  !q := !q + 1 // : Unit$^\lstcm{\qbot}$ $\JUSTU{\texttt{q}}$
}
free(q) // : Unit$^\lstcm{\qbot}$ $\JUSTK{\texttt{q}}$
defer_access() // : $\JUSTK{\texttt{q}} \NOTEFFSEQ \JUSTU{\texttt{q}}$
// $\lsterrcm{\text{Static Error: use after free}}$
    \end{lstlisting}
    \caption{RT (\textbf{this work}) permits shared-reference programs, so effects track the destructive operation.}
    \label{fig:intro-shared-rt}
  \end{subfigure}
  \caption{Destructive operations on shared resources.}
  \label{fig:intro-shared-destruction}
\end{figure*}
 \subsection{Disabling Aliases: Destructive Operations on Shared Resources}
\label{sec:intro-shared-destruction}
\footnotetext{Because our system natively supports full-fidelity modeling of
closures and higher-order control flow, pairs and tuples can be Church encoded
and need not be primitive to the core calculus.}

The \texttt{freeze} closure in \Cref{fig:intro-counter-freeze} performs an
action that most type systems forbid outright: it \emph{deallocates} state that
remains reachable through other closures. Why is this difficult to support? As
shown in \Cref{fig:intro-capture-vs-call}, Rust must move the counter into
\texttt{freeze}, because it cannot permit multiple live \emph{mutable} references
to the same state; shared mutable aliasing is forbidden by construction.
Nevertheless, sharing mutable state across sibling closures is a ubiquitous and
valuable pattern. Rust recovers this capability via runtime reference counting
(\eg \texttt{Rc}). However, once a
resource is managed by an \texttt{Rc}, it is reclaimed only when its final owner
is dropped, forcing the programmer to surrender manual deallocation. 
Consequently, existing ownership systems force a strict dichotomy:
developers must choose between shared mutability and deterministic reclamation,
but cannot have both.

We now turn to the essential mechanism in our system that resolves this tension:
a kill effect applied through one name must \emph{disable every alias} that
reaches the resource. This ensures a program can maintain shared access to a
resource \emph{and} still reclaim it explicitly. While deallocation is a typical
use case, this same kill effect can be applied in revoking non-memory resources,
such as file handles, locks, and capabilities. In alias-heavy APIs, one name may
destroy or detach a resource while another name still denotes it. 

Because the deallocated alias is often captured within a callback or
handler, a use-after-free error can be \emph{latent}---captured by a
closure before the kill effect is actually triggered. Consequently, a
robust safety guarantee must account for both immediate accesses through an
active alias \emph{and} deferred accesses invoked later via a closure.

\Cref{fig:intro-shared-destruction} contrasts three paradigms.  
Unsafe APIs (\Cref{fig:intro-shared-ir}) let one alias delete a resource
and leave the others dangling, risking runtime errors. Rust (\Cref{fig:intro-shared-rust}) closes that
hole by making sharing and deallocation mutually exclusive: with \texttt{Rc},
several holders may share and mutate the resource, but none may free it
explicitly. Our effectful operator \tfree\ instead induces a \textit{kill}
effect, statically forbidding later dereference or assignment through the killed
reachability set, so a program can share a resource \emph{and} deallocate it
manually (\Cref{fig:intro-shared-rt}).

Because the kill effect applies equally to non-memory resources, this mechanism
naturally encompasses typestate-style capability
revocation~\citep{jia_typestate_2025}. Finally, unlike
region- and arena-based approaches~\citep{he_when_2025}---which reclaim memory
at lexical boundaries and may keep unused references live until the entire arena
is collected---our flow-sensitive effect system can track effects through
higher-order control flow and free a reference precisely at the branch where it
becomes unused.

\subsection{Deriving Exclusivity: Uniqueness for Recursively Isolated Lists}
\label{sec:intro-uniqueness}

\begin{figure}[tb]
  \centering
  \begin{subfigure}[b]{0.5\textwidth}
    \centering
\begin{lstlisting}[basicstyle=\ttfamily\scriptsize,breaklines=true]
// Destructive merge of two unique linked lists
val l1 = uref(cons(1, uref(cons(3,
           uref(cons(7, uref(nil)))))))
val l2 = uref(cons(2, uref(cons(5,
           uref(cons(8, uref(nil)))))))
val l3 = merge(l1, l2)  $\lstcm{\leftarrow \text{ moves \texttt{l1}, \texttt{l2}}}$
                        // : $\lstcm{\UKX{\texttt{l1,l2}}{\texttt{l1,l2}}}$ 
// use(l1)              $\lsterrcm{\leftarrow \text{ Error: \texttt{l1} moved}}$
\end{lstlisting}
    \caption{A destructive \texttt{merge} of two unique linked lists. The
    call's effect \emph{moves} (\ie both \textit{use}s and \textit{kill}s)
    \texttt{l1} and \texttt{l2}.
    }
    \label{fig:list-code}
  \end{subfigure}
  \hfill
  \begin{subfigure}[b]{0.49\textwidth}
    \centering
    \resizebox{\linewidth}{!}{%
    \begin{tikzpicture}[
        font=\tiny,
        list node/.style={draw, minimum width=0.4cm, minimum height=0.32cm, inner sep=1pt},
        l1 node/.style={list node, draw=orange, fill=orange!20},
        l2 node/.style={list node, rounded corners, draw=blue, fill=blue!20},
        l3 sharp/.style={list node, draw=purple, fill=purple!20},
        l3 round/.style={list node, rounded corners, draw=purple, fill=purple!20},
        own/.style={draw=#1, dashed, rounded corners},
        veil/.style={fill=black!45, opacity=0.35, rounded corners},
        rootarrow/.style={->, >=stealth, thick, shorten >=1pt},
      ]
      \node[l1 node] (a1) at (0,0) {1};
      \node[l1 node, right=0.22cm of a1] (a2) {3};
      \node[l1 node, right=0.22cm of a2] (a3) {7};
      \draw[->, orange] (a1) -- (a2);
      \draw[->, orange] (a2) -- (a3);

      \node[l2 node, right=1.5cm of a3] (b1) {2};
      \node[l2 node, right=0.22cm of b1] (b2) {5};
      \node[l2 node, right=0.22cm of b2] (b3) {8};
      \draw[->, blue] (b1) -- (b2);
      \draw[->, blue] (b2) -- (b3);

      \node[veil, fit=(a1)(a3), inner sep=6pt] {};
      \node[veil, fit=(b1)(b3), inner sep=6pt] {};

      \node[own=orange, fit=(a2)(a3), inner sep=2pt] {};
      \node[own=orange, fit=(a1)(a3), inner sep=6pt] (r1) {};
      \node[own=blue,   fit=(b2)(b3), inner sep=2pt] {};
      \node[own=blue,   fit=(b1)(b3), inner sep=6pt] (r2) {};

      \node[text=orange, left=0.42cm of r1.west] (l1lab) {\texttt{l1}};
      \draw[rootarrow, orange] (l1lab) -- (a1.west);
      \node[text=blue, left=0.42cm of r2.west] (l2lab) {\texttt{l2}};
      \draw[rootarrow, blue] (l2lab) -- (b1.west);

      \node[l3 sharp] (m1) at (0.7cm,-1.2cm) {1};
      \node[l3 round, right=0.22cm of m1] (m2) {2};
      \node[l3 sharp, right=0.22cm of m2] (m3) {3};
      \node[l3 round, right=0.22cm of m3] (m4) {5};
      \node[l3 sharp, right=0.22cm of m4] (m5) {7};
      \node[l3 round, right=0.22cm of m5] (m6) {8};
      \draw[->, purple] (m1) -- (m2);
      \draw[->, purple] (m2) -- (m3);
      \draw[->, purple] (m3) -- (m4);
      \draw[->, purple] (m4) -- (m5);
      \draw[->, purple] (m5) -- (m6);

      \coordinate (g5) at ($(m4.east)!0.5!(m5.west)$);
      \coordinate (g4) at ($(m3.east)!0.5!(m4.west)$);
      \coordinate (g3) at ($(m2.east)!0.5!(m3.west)$);
      \coordinate (g2) at ($(m1.east)!0.5!(m2.west)$);
      \draw[own=purple] ($(g5 |- m1.north)+(0,0.05cm)$) rectangle ($(m6.south east)+(0.05cm,-0.05cm)$);
      \draw[own=purple] ($(g4 |- m1.north)+(0,0.12cm)$) rectangle ($(m6.south east)+(0.12cm,-0.12cm)$);
      \draw[own=purple] ($(g3 |- m1.north)+(0,0.19cm)$) rectangle ($(m6.south east)+(0.19cm,-0.19cm)$);
      \draw[own=purple] ($(g2 |- m1.north)+(0,0.26cm)$) rectangle ($(m6.south east)+(0.26cm,-0.26cm)$);
      \draw[own=purple] ($(m1.north west)+(-0.10cm,0.33cm)$) rectangle ($(m6.south east)+(0.33cm,-0.33cm)$);

      \node[text=purple] (l3lab) at ($(m1.west)-(0.62cm,0)$) {\texttt{l3}};
      \draw[rootarrow, purple] (l3lab) -- (m1.west);

      \coordinate (owner) at ($(m6.east)+(0.33cm,0)$);
      \draw[rootarrow, gray, dashed] ($(owner)+(0.75cm,-0.28cm)$) -- (owner);
      \draw[red, very thick] ($(owner)+(-0.06cm,-0.06cm)$) -- ($(owner)+(0.06cm,0.06cm)$);
      \draw[red, very thick] ($(owner)+(-0.06cm,0.06cm)$) -- ($(owner)+(0.06cm,-0.06cm)$);
    \end{tikzpicture}%
    }
    \caption{Ownership transfer: \textcolor{orange}{\texttt{l1}} and
    \textcolor{blue}{\texttt{l2}} are consumed; \textcolor{purple}{\texttt{l3}}
    becomes the only owner of the merged list. Dashed boxes enclose each
    ownership domain; node shapes (sharp vs.\ rounded) trace element provenance.}
    \label{fig:list-diagram}
  \end{subfigure}
  \caption{Ownership transfer during a destructive \texttt{merge} of two unique
  linked lists.  See \Cref{sec:case-merge} for the full implementation of \texttt{merge}.}
  \label{fig:list-ownership}
\end{figure}
 
The first two ingredients allow closures to safely share and eventually consume
aliased state without enforcing rigid global invariants. However, exclusivity
remains fundamentally important for safe resource manipulation—so how does our
system recover it when a program actually needs it? Rather than falling back on
built-in uniqueness primitives that clash with closures, our system
\emph{derives} exclusivity on demand. Because the effect system already accounts
for every name that can reach a resource, uniqueness emerges organically as a
mode of use from reachability and flow-sensitive \textit{use}/\textit{kill}
effects, preserving full compatibility with higher-order control flow.

Modeling uniqueness in this way specifically addresses the strict constraints of
a typed higher-order $\lambda$-calculus. In first-order or statement-level
settings, systems can freely propagate precise and intricate store abstractions
sequentially from one statement to the
next~\citep{DBLP:conf/pldi/MilanoTM22,clebsch_deny_2015,Proc.ACMProgram.Lang./ArvidssonESSJM23}.
In contrast, a higher-order system must represent state in compact,
human-readable types capable of traversing through escaping closures. By
transforming uniqueness from a rigid annotation into an emergent property, we
preserve full compatibility with higher-order control flow while freeing the
user from maintaining global uniqueness constraints.

Crucially, derived exclusivity is not confined to single reference cells. Built
from the same alias tracking that already scales to higher-order functions and
nested structures, it extends to recursively \emph{isolated} data structures
such as linked lists, where each node uniquely owns its tail thus ensuring
exclusive access to the data. 

\Cref{fig:list-ownership} makes this concrete by demonstrating a destructive
\texttt{merge} operation over two unique linked lists: the inputs \texttt{l1}
and \texttt{l2} are consumed, and the result \texttt{l3} becomes the sole
live handle to the merged data. This exclusivity is not declared anywhere in
\texttt{merge}'s signature; it falls out entirely from the effect tracking that
both inputs are killed by the call, so no alias to either survives it.
The specific operators that realize this feature are deferred to
\Cref{sec:extension} and \Cref{sec:extension-lists}.

\subsection{Contributions and Organization}

Overall, this paper makes the following contributions:

\begin{itemize}

  \item \textbf{Ownership transfer in the presence of higher-order control
  flow.} We enable ownership transfer via a flow-sensitive effect system that
  tracks \textit{use} and \textit{kill} effects. Rather than treating closures
  as owners, our system carries effects latently on function types and induces
  them upon application.  This allows consumption to be deferred across
  higher-order control flow while statically preventing use-after-kill errors.

  \item \textbf{Safe deallocation in the presence of mutable references.} The
  same \textit{use}/\textit{kill} effects give no-use-after-kill
  guarantees for destructive operations on resources---deallocation as well as
  disabling of files or capabilities---even in the presence of mutable sharing.

  \item \textbf{Uniqueness derived from reachability tracking and
  flow-sensitive effects.} We demonstrate that with reachability tracking and
  flow-sensitive effects, unique references and recursively isolated lists can
  be expressed as derived properties.

  \item \textbf{A diverse suite of case studies.} We present a comprehensive set
  of case studies---a destructive merge on unique lists, a freezable counter,
  Rust-style swap on \texttt{Box}, recursive sum on unique lists, and use-once
  continuations---demonstrating expressiveness.

  \item \textbf{Mechanized metatheory.} We prove soundness (progress,
  preservation, multi-step preservation, and preservation under parallel
  reduction) and mechanize the key results in Rocq.

\end{itemize}

\paragraph{Overview} 

The rest of the paper develops the design, metatheory, and
applications of this effect system. \Cref{sec:motivation} revisits
reachability types and introduces motivating examples that illustrate how
reachability and flow-sensitive effects capture common idioms such as resource
destruction, ownership transfer on individual references, and unique
references derived from these effects.
\Cref{sec:formal} formally presents the \maybelang calculus, defining its
syntax, typing/subtyping rules, operational semantics, and meta-theoretical
results, with a particular emphasis on flow-sensitive effects.
\Cref{sec:extension} and \Cref{sec:extension-lists} extend the core system with
unique references, atomic swap, and linked lists. \Cref{sec:casestudy} presents
several case studies evaluating \maybelang's expressivity. Finally,
\Cref{sec:limitation} explores future directions and \Cref{sec:related} surveys
related work. All formal results are mechanized in Rocq; we redact links for
anonymous review and submit them as anonymized supplementary material.

To our knowledge, we are the first to formulate ownership transfer in a
\textit{higher-order setting} while \emph{retaining shared mutable aliasing},
and \emph{deriving} uniqueness as a mode of use. %
\section{Reachability Types (RT)} \label{sec:motivation}

In this section, we revisit the key components of reachability types as
introduced by \citet{DBLP:journals/pacmpl/WeiBJBR24} and extend them, topic by
topic, with the flow-sensitive effect layer that this paper adds. We begin with
reachability qualifiers and contextual freshness
(\Cref{sec:motivation-qualifiers}), then introduce the primitive
\textit{use}/\textit{kill} effect operators (\Cref{sec:motivation-effects}). We
then contrast idempotent kill with non-idempotent move
(\Cref{sec:motivation-idempotent}), present unique references, whose uniqueness
is \emph{derived} from the effect system rather than taken as a primitive
(\Cref{sec:motivation-unique}), and finally discuss their behavior under
higher-order functions through latent effects and fresh application
(\Cref{sec:motivation-latent}). We formalize these ideas in the \maybelang
calculus in \Cref{sec:formal}.

\subsection{Reachability Qualifiers and Contextual Freshness}
\label{sec:motivation-qualifiers}

Types in RT are of the form $\ty[p]{T}$, where $p$ is a \textit{reachability
qualifier}, indicating the set of runtime locations that may be reached from the
result of an expression. Statically, these locations are approximated by
variables and the optional freshness marker $\QFresh$, indicating a fresh,
unnamed resource.

In the following example, evaluating the expression @new Ref(0)@ results in a
\emph{fresh} value: it is not yet bound to a name, but must be tracked. The
typing context \lstinline|$\lstcm{\text{[ counter: Ref[Int]}^{\QFresh} ]}$|
following a reverse turnstile "$\dashv$" means @counter@ reaches a fresh value:

\begin{lstlisting}
  val counter = new Ref(0)        // : Ref[Int]$^\lstcm{\trackvar{counter}}$ $\lstcm{\dashv}$ [ counter: Ref[Int]$^\lstcm{\QFresh}$ ]
  val counter2 = counter          // : Ref[Int]$^\lstcm{\trackvar{counter2}}$ $\lstcm{\dashv}$ [ counter2: Ref[Int]$^\lstcm{\trackvar{counter}}$, counter: Ref[Int]$^\lstcm{\QFresh}$ ]
\end{lstlisting}

RT keeps reachability sets minimal (\eg, @counter@ tracks exactly itself); when an
alias @counter2@ is created, RT assigns it the one-step reachability set @counter@.
We can retrieve the complete reachability set by computing its transitive closure
with respect to the typing context~\cite{DBLP:journals/pacmpl/WeiBJBR24}.

Functions also track reachability: their reachability qualifier includes all
\textit{captured variables}. In the following example, function @inc@ captures
the free variable @counter@:

\begin{lstlisting}
  def inc(n: Int) = { counter := !counter + n } // : inc: (Int => Unit)$\lstcm{\trackvar{counter}}$ $\lstcm{\dashv}$ [ counter: Ref[Int]$^\lstcm{\QFresh}$ ]
\end{lstlisting}

The presence of the freshness marker \QFresh\ in a function argument's qualifier
indicates that the argument may only reach \textit{unobservable} resources,
meaning that its reachable locations must remain separate from those of the
function. Thus, function applications must satisfy the \textit{separation
constraint}, requiring that the argument's reachability qualifier is disjoint
from that of the function.

\begin{lstlisting}
  def id(x: T$^\QFresh$): T$\trackvar{x}$ = x                                       // : ((x: T$^\lstcm{\QFresh}$) => T$^\lstcm{\trackset{x}}$)$^\lstcm{\qbot}$
\end{lstlisting}

\noindent The type means that function @id@ cannot capture anything from its context, and
it accepts arguments that may reach unobservable resources. A function application that violates
this separation constraint results in a type error:

\begin{lstlisting}
  def update(x: Ref[Int]$^\QFresh$): Unit = { counter := !(x) + 1 }   // : ((x: Ref[Int]$^\lstcm{\QFresh}$) => Unit)$^\lstcm{\trackvar{counter}}$
  update(counter) // $\lsterrcm{\text{Error! variable counter overlaps with function update}}$
\end{lstlisting}

\noindent The above function application violates the
separation constraint: the passing argument @counter@ overlaps with function @update@.

\subsection{Flow-Sensitive Effects: Use and Kill}
\label{sec:motivation-effects}

Without flow-sensitive effects, RT can already model a rich set of resource usage
patterns, using regions to model lexically scoped
lifetimes~\Citep{he_when_2025}. However, regions are not expressive enough to
model patterns that involve non-lexical lifetimes, such as explicit deallocation
and ownership transfer. To support such patterns, we extend RT with a
flow-sensitive effect system that records how tracked resources are accessed as
the program executes. Intuitively, we distinguish two families of effects: a
\textit{use} effect marks that a resource is observed or mutated, whereas a
\textit{kill} effect records that a resource becomes unavailable for any
subsequent use.

This system is inspired by extensions proposed in prior
works~\Citep{DBLP:journals/pacmpl/BaoWBJHR21,DBLP:journals/pacmpl/WeiBJBR24,Proc.ACMProgram.Lang./DengSSY25}.
We write effects in record syntax \(\UKX{\alpha}{\beta}\) where \(\alpha\) and
\(\beta\) are finite sets of variables annotated with labels \(\USE\)
(\emph{used} variables) and \(\KILL\) (\emph{killed} variables). Empty effect
components are omitted for readability (see \Cref{fig:maybe-syntax}). Two
effects can be combined using the flow-sensitive effect sequencing operator
\EFFSEQ. Function types carry both reachability qualifiers and latent effects,
and the typing rules sequence effects when composing computations. This
combination gives us a uniform view: qualifiers continue to describe
\textit{what} resources a value can reach, while effects describe \textit{how}
those resources are accessed as the program executes.

\paragraph{Use effects for reads and writes.}
Whenever an effectful operation accesses a tracked resource, the effect system
records that access. Reads and writes therefore induce use effects on the
resource's reachability set. In the following snippet the assignment does not
just update the store; it also registers a \(\USE\) effect that must be
accounted for at the call site:

\begin{lstlisting}
  val cell = new Ref(elm) // : Ref[T$^\lstcm{\trackvar{elm}}$]$^\lstcm{\trackvar{cell}}$
  cell := ... // : $\JUSTU{\texttt{cell}}$ $\lstcm{\leftarrow \text{using \texttt{cell} but not using \texttt{elm}}}$
\end{lstlisting}

\paragraph{Kill effects for deallocation.}
Explicit deallocation is tracked via \textit{kill} effects. The \tfree\
construct deallocates a location and tracks it in the set of killed resources:

\begin{lstlisting}
  free(cell) // : $\JUSTK{\texttt{cell}}$ $\lstcm{\leftarrow \text{killing cell}}$
\end{lstlisting}

Consequently, the static effect sequencing operator (\EFFSEQ, see
\Cref{fig:effect-ops}) must ensure that no later computation can accidentally
rely on this dead value. The effect trace \(\JUSTK{\texttt{cell}}\) witnesses
that the subsequent computation must not use the reference after the
deallocation point.

\paragraph{Main invariant: no use-after-kill.}
The effect system ultimately enforces a global invariant on
program traces: once a variable is killed, it cannot be used again:

\begin{lstlisting}
  !cell // : $\JUSTK{\texttt{cell}} \NOTEFFSEQ \JUSTU{\texttt{cell}}$ $\lsterrcm{\text{Error! Attempting to use killed variable cell!}}$
\end{lstlisting}

This invariant is enforced through the type
system, which tracks effects in function types and checks effect sequencing at
function call sites.

\paragraph{Use vs.\ mention.}\label{sec:motivation-use-vs-mention}
We separate \emph{using} a resource from merely \textit{mentioning} it in a
type.  A mere mention does not induce an effect; therefore, expressions that only mention resources
remain pure.~\footnote{Programs in the systems presented in this work always terminate~\cite{bao_modeling_2025}. Thus, an expression merely mentioning resources in this work is pure~\cite{bao2025typeabilityeffectsystems}.} Moreover, merely \emph{mentioning} a killed variable is allowed and does not
induce any effect:

\begin{lstlisting}
  def size(c: Ref[T$\trackvar{elm}$]$^\QFresh$) = 0 // : (c: Ref[T$\lstcm{\trackvar{elm}}$]$\lstcm{\trackvar{\QFresh}}$) => Int$^\lstcm{\qbot}$ $\PURE$
  size(cell) // : Int$^\lstcm{\qbot}$ $\PURE$ $\lstcm{\leftarrow \text{ok: mere mention of \texttt{cell}}}$
\end{lstlisting}

\paragraph{Precise effect tracking.} Effect annotations overapproximate the
resources an operation actually manipulates, but precision is equally important.
Building on the one-step store reachability of
\Citet{Proc.ACMProgram.Lang./DengSSY25}---where reference qualifiers track
individual locations rather than transitively reachable ones---our effects
precisely track used and deallocated locations: reassigning @cell@ marks @cell@
as \textit{used} without propagating the use effect to the enclosed resource
@elm@:

\begin{lstlisting}
  cell := ...             // : $\JUSTU{\texttt{cell}}$                           $\lstcm{\leftarrow \text{using cell but not using elm}}$
  free(cell)              // : $\JUSTU{\texttt{cell}} \EFFSEQ \JUSTK{\texttt{cell}}$                           $\lstcm{\leftarrow \text{killing cell}}$
  use(elm)                // : Unit$^\lstcm{\qbot}$ $\JUSTU{\texttt{elm}}$                      $\lstcm{\leftarrow \text{using elm, okay}}$
\end{lstlisting}

\subsection{Idempotent Kill and Non-Idempotent Move}
\label{sec:motivation-idempotent}

\paragraph{Idempotent kill effects.}\label{sec:motivation-effects-idempotent}
Kill effects are \textit{idempotent}: freeing a resource more than once is allowed
and behaves like freeing it once, while any subsequent \emph{use} of the killed
resource still fails (see \Cref{sec:formal-dynamic}).

\paragraph{Non-idempotent, explicit, Rust-style move semantics.}
On the other hand, we introduce a non-idempotent \tmove\ construct that
simultaneously induces a \textit{kill} and \textit{use} effect. We also refer to
this combination of effects as a \textit{move} effect. The fact that
\textit{move} effect is non-idempotent, as opposed to \textit{kill}, is
precisely because of the \textit{use} effect that is additionally induced, so
that if a resource is already killed or moved, it cannot be \textit{moved}
again.

Additionally, \tmove\ returns the argument as a fresh resource. In the example
below, reference cell @r@ is \textit{moved}, and its ownership is transferred to
@s@ via the returned fresh value:

\begin{lstlisting}
  val r = new Ref(0)    // : Ref[Int]$^\lstcm{\trackvar{r}}$
  val s = move r        // : $\JUSTM{\texttt{r}}$ $\lstcm{\dashv}$ [ s: Ref[Int]$^\lstcm{\QFresh}$ ]
\end{lstlisting}

After the move, @r@ becomes unusable, and any further use of @r@ is a type error:

\begin{lstlisting}
  r := 5                // : $\lsterrcm{\text{Error! Cannot use moved resource r}}$
\end{lstlisting}

However, @s@ now owns the resource originally owned by @r@, and can be used
safely:

\begin{lstlisting}
  s := 41               // : $\JUSTU{\texttt{s}}$
  free(s)               // : $\JUSTU{\texttt{s}} \EFFSEQ \JUSTK{\texttt{s}}$
\end{lstlisting}

We describe the formal semantics of \tfree\ and \tmove\ in \Cref{sec:formal}.

\paragraph{Why a primitive move?}
A primitive \tmove\ can, in principle, be encoded: a move is a \texttt{borrow} whose
block is the \emph{entire continuation} of the move point, so that control
never returns to a scope where the old name is live. The \texttt{borrow}
combinator introduces a fresh, separate name for its argument and prevents
the block from mentioning the old one:

\begin{lstlisting}
  def borrow[A$^\QFresh$,B$^\QFresh$](x: A$^\QFresh$) (block: (A$\trackvar{x}$ => B$^{\QFresh\texttt{,x}}$)$^\qbot$): B$^{\QFresh\texttt{,x}}$ = block(x)
\end{lstlisting}

\noindent A selective CPS transformation---or \texttt{shift}/\texttt{reset} in
direct style~\citep{DBLP:conf/lfp/DanvyF90,DBLP:conf/popl/Filinski94}---reifies
that continuation as the block:

\begin{lstlisting}
  val r = new Ref(0)             // : Ref[Int]$^\lstcm{\trackvar{r}}$
  reset { val s = shift k { borrow(r)(k) }
          ... }                  // $\lstcm{\leftarrow \text{use borrowed resource as \texttt{s}; \texttt{k} cannot mention \texttt{r}}}$
\end{lstlisting}

\noindent We nonetheless retain a direct-style \tmove. While an encoding can
express transfer, it requires a global program transformation and does not
perform actual reclamation---only the effect semantics actually deallocate
(\Cref{sec:formal-dynamic}). 

\subsection{Unique References, Derived Rather Than Primitive}
\label{sec:motivation-unique}

Beyond freeing cells and basic move semantics, the effect system also
expresses unique references---and, crucially,
their uniqueness is \emph{derived} from alias tracking and the
\textit{use}/\textit{kill} effects rather than taken as a primitive
\texttt{iso}-style
capability~\citep{clebsch_deny_2015,DBLP:conf/pldi/MilanoTM22}. The \turef\ constructor introduces a fresh, nested
reference whose qualifier guarantees exclusive access to the underlying referent,
inducing a move effect for the resource that was moved in. In the code snippet
below, @tmp@ is moved into the unique reference @ur@, which disables any further
access to itself, transferring its ownership to @ur@:

\begin{lstlisting}
  val tmp = new Ref(0)           // : Ref[Int]$^\lstcm{\trackvar{tmp}}$
  val ur = uref(tmp)             // : Ref[Ref[Int]$^\lstcm{\QFresh}$]$^\lstcm{\trackvar{ur}}$ $\JUSTM{\texttt{tmp}}$
\end{lstlisting}

Once a unique reference is created, \tswap\ can exchange the payload by
consuming/moving another resource while returning its original referent as a
fresh resource:

\begin{lstlisting}
  val r = new Ref(41)            // : Ref[Int]$^\lstcm{\trackvar{r}}$
  val fr = swap(ur, r)           // : Ref[Int]$^\lstcm{\QFresh}$ $\UKX{\texttt{ur,r}}{\texttt{r}}$ $\lstcm{\leftarrow \text{ur cell now holds 41}}$
  !fr                            // : Int$^\lstcm{\qbot}$ $\lstcm{\leftarrow \text{0}}$
\end{lstlisting}

In the above code snippet, @r@ is a newly created resource that is consumed when
being swapped into @ur@. The \tswap\ operator returns a fresh reference for its
original payload. Upon swapping, the unique reference @ur@ is only \textit{used}
but not \textit{killed}. We describe unique references and swapping in full
detail in \Cref{sec:extension}.

\subsection{Latent Effects and Fresh Application}
\label{sec:motivation-latent}

The constructs so far transfer ownership of a single named reference. The same
\textit{use}/\textit{kill} discipline extends to ownership transfer \emph{through
a higher-order value}, because function types carry \emph{latent effects}: a
closure records, in its type, the \textit{use} and \textit{kill} effects it will
induce when applied. Flow-sensitive reasoning must therefore live in the function
type itself, and a fresh application returns the remaining access path. Fresh
application also constrains the callee: a fresh argument must not be both
\emph{returned} and \emph{killed} by the function, which would otherwise leak a
dangling access path that escapes tracking (see \Cref{sec:formal-app}).

Consider the freezable counter of \Cref{fig:intro-counter-freeze}. The value
\texttt{counter(42)} bundles two closures over one shared state cell \texttt{p}:
\texttt{incr} carries a latent \textit{use} effect \(\JUSTU{\texttt{p}}\); and
\texttt{freeze} carries a latent \textit{use}/\textit{kill} effect
\(\UKX{\texttt{p}}{\texttt{p}}\), so that invoking it both \emph{uses} and
\emph{kills} \texttt{p}. Effect sequencing then rejects any later call that uses
\texttt{p} after \texttt{freeze} has killed it---including a call through the
sibling closure \texttt{incr}. Concretely, the initial \texttt{frz()} is well
typed with effect \(\UKX{\texttt{p}}{\texttt{p}}\), but a subsequent
\texttt{incr()} is rejected because \(\JUSTK{\texttt{p}} \NOTEFFSEQ
\JUSTU{\texttt{p}}\). Ownership of the shared state has been consumed
\emph{through} the \texttt{freeze} closure, not by moving any closure value, and
the sound rejection of \texttt{incr} after \texttt{frz} follows from the same
effect sequencing that rules out use-after-free.

\section{\maybelang: Reachability Types with Destructive Effects}\label{sec:formal}

\begin{figure}[t]\footnotesize

\judgement{Syntax}{\BOX{\maybelang}}\small\vspace{-10pt}
\[\begin{array}{l@{\qquad}l@{\qquad}l@{\qquad}l}
    x,y,z     & \in & \Var                                                                                    & \text{Variables}           \\
    f,g,h     & \in & \Var                                                                                    & \text{Function Variables}  \\
    X         & \in & \Var                                                                                    & \text{Type Variables}      \\ [.5ex]
    S,T,U,V   & ::= & \TUnit \mid f(x: \ty{Q}) \to \ty{R} \mid \mr{}{Q}{}                                     &                            \\
              &     & \mid \TTop \mid X \mid \forall f(\ty[x]{X} <: Q). Q                                     & \text{Types}               \\
    B         & ::= & \TUnit                                                                                  & \text{Base Types}          \\
    t,t_1,t_2 & ::= & c \mid x \mid \lambda f(x).t \mid t_1~t_2 \mid \tref~t \mid\ !~t \mid t_1 \coloneqq t_2 &                            \\
              &     & \mid \TLam{X}{x}{T}{q}{t} \mid \TApp{t}{Q}{} \mid \ext*{\tfree~t} \mid \ext*{\tmove~t}    & \text{Terms}               \\ [.5ex]
    p,q,r,w   & \in & \mathcal{P}_{\mathsf{fin}}(\Var \uplus \{ \QFresh \})                                   & \text{Type Qualifiers}     \\
    \alpha,\beta,\gamma 
              & \in & \mathcal{P}_{\mathsf{fin}}(\Var)                                                        & \text{Effect Qualifiers}   \\ [.5ex]
    O,P,Q,R   & ::= & \ty[q]{T}                                                                               & \text{Qualified Types}     \\
    \ext{\EPS,\EPS[1],\EPS[2],\EPS[3]} 
              & ::= & \ext{\UKX{\alpha}{\alpha}}                                                              & \text{Effects}             \\ [.5ex]
    \flt      & \in & \mathcal{P}_{\mathsf{fin}}(\Var)                                                        & \text{Observations}        \\
    \Gamma    & ::= & \varnothing\mid \Gamma, x : Q   \mid \Gamma, \ty[x]{X} <: Q                             & \text{Typing Environments} \\
\end{array}\]

\footnotesize
\judgement{Type, Qualifier, and Effect Notations}\small\vspace{-2pt}
\[\begin{array}{l@{\qquad}l@{\qquad}l@{\qquad\qquad\qquad\qquad}l}
    \ext{\PURE} & := & \ext{\UKX{\varnothing}{\varnothing}} & \text{Empty Effect}              \\
    p,q         & := & p \qlub q                            & \text{Qualifier Union}           \\
    x           & := & \{x\}                                & \text{Single Variable Qualifier} \\
    \QFresh     & := & \{\QFresh\}                          & \text{Single Fresh Qualifier}    \\
\end{array}\]

\caption{The syntax of \maybelang. \ext{\text{Effects qualifiers, notations}}
and \ext*{\text{effectful operations}} are highlighted, representing the core
additions of this work over prior reachability type
systems.}\label{fig:maybe-syntax}

\vspace{-3ex}
\end{figure}

In this section, we formally present \maybelang.  \Cref{sec:formal-syntax}
introduces variables, terms, ordinary and qualified types, effects, and
environments; \Cref{sec:formal-static} formalizes the typing/effect judgments
for expressions; \Cref{sec:formal-subtyping} specifies subtyping for types,
qualifiers, and effects; \Cref{sec:formal-dynamic} presents the small-step
operational semantics, particularly highlighting the treatment of store
deallocation and ownership transfer; finally, \Cref{sec:formal-meta} describes
the meta-theoretical results including preservation (\Cref{thm:preservation}),
multi-step type safety (\Cref{lem:multi-step-type-safety}), and parallel
progress and preservation (\Cref{cor:parallel-progress}).

\subsection{Syntax}\label{sec:formal-syntax}

\Cref{fig:maybe-syntax} presents the surface syntax for \maybelang. It extends
\Citet{Proc.ACMProgram.Lang./DengSSY25}'s reachability types system with constructs for
explicit deallocation and ownership transfer, along with a flow-sensitive effect
system.

\subsubsection{Qualifiers} Type qualifiers \(p,q,r,w \in
\mathcal{P}_{\mathsf{fin}}(\Var \uplus \{\QFresh\})\) are finite sets of names,
optionally including the distinct freshness marker \(\QFresh\).  Effect
qualifiers \(\alpha,\beta,\gamma\) range over the same carrier set but do not
include the fresh marker \(\QFresh\) (see \Cref{sec:formal-app}). Effect
qualifiers are parameterized components of an effect. For readability we often
write qualifiers as comma-separated lists of names rather than set brackets (see
\Cref{fig:maybe-syntax}).

\subsubsection{Effects} 

We use \(\EPS,\EPS[1],\EPS[2],\EPS[3]\) to represent effect variables. Effects
are records carrying two qualifiers, representing its \textit{use} component
(\FX{\USE}) and \textit{kill} component (\FX{\KILL}). Effect
\(\UKX{\alpha}{\alpha}\) is parameterized by the \textit{use} and the
\textit{kill} component respectively. When one component is empty, we may omit
the empty component by writing \(\JUSTU{\alpha}\) or \(\JUSTK{\alpha}\).

\subsubsection{Types and Qualified Types}

We separate ordinary types \(T\) from qualified types \(Q\).  A qualified type
has the form \(Q \equiv \ty[q]{T}\), where the qualifier \(q\) tracks a finite
set of names relevant to \(T\) (e.g., potential aliases/reachable variables).
We use \(S,T,U,V\) to range over ordinary types and \(O,P,Q,R\) over qualified
types.

\paragraph{Base Types.} Base types do not carry qualifiers as a sub-component,
and in our system, they are \(\TUnit\), the unit type, and \(\TTop\), the top
type.

\paragraph{Dependent Function Types.} Dependent function types are of the form
\(f(x: \ty[q]{T}) \to^{\EPS} \ty[p]{S}\), where the codomain \(\ty[p]{S}\) may
depend on the argument \(x\) and the self-reference \(f\).  Function types also
carry a latent effect $\EPS$.  The latent effect may depend on both the function
self-reference \(f\) and argument \(x\), just as the return type-and-qualifier
\(\ty[p]{S}\) do. 

\paragraph{Reference Types.} Reference types have the form \(\mr{}{T}{q}\),
 where \(T\) is the referent type and \(q\) is the referent qualifier that
 tracks the reachable variables through dereferencing.

\paragraph{Universal Types.} Universal types \(\forall\,f(\ty[x]{X} <: Q).\, Q\)
quantify over a type variable \(X\) with a qualifier \(x\) under a qualified
upper bound \(Q\); the self reference \(f\) is brought into scope so that both
types and qualifiers in the body may mention the universal type as a whole.

\paragraph{Terms.} Constants \(c\) inhabit base types \(B ::= \TUnit\).
\(\lambda f(x).t\) is a recursive function (binding self \(f\) and parameter
\(x\) in \(t\)); application has the form of \(t_1~t_2\). Additionally, we have
reference manipulating terms \(\tref~t\), \({!}\,t\), and \(t_1 \coloneqq t_2\),
which perform reference allocation, dereference, and assignment respectively.
\(\TLam{X}{x}{T}{q}{t}\) is a bounded type abstraction over a type variable
\(X\) with qualifier \(x\), \(\TApp{t}{Q}{}\) instantiates with a qualified type
argument \(Q\). The forms \(\tfree~t\) and \(\tmove~t\) are primitive effectful
operations whose effects are tracked in the static semantics. We extend the
system with two additional effectful operators \turef\ and \tswap\ in
\Cref{sec:extension}.

\subsection{Static Typing}\label{sec:formal-static}

\begin{figure}[t]\footnotesize

  \judgement{Term Typing}{\BOX{\strut\G[\flt] \ts t : \ty{Q}\ \EPS}}\\[1ex]
  \begin{minipage}[t]{.25\linewidth}\small\vspace{0pt}
    \typerule{t-var}{
      y : \ty[q]{T} \in \G\quad y \in \flt
    }{
      \G[\flt] \ts y : \ty[y]{T}\ \PURE
    }
  \end{minipage}%
  \begin{minipage}[t]{.25\linewidth}\small\vspace{0pt}
    \typerule{t-cst}{
      c \in B
    }{
      \G[\flt] \ts c : \ty[\qbot]{B}\ \PURE
    }
  \end{minipage}%
  \begin{minipage}[t]{.5\linewidth}\small\vspace{0pt}
    \typerule{t-sub}{
      \G[\flt]\ts t : \ty{Q}\ \EPS[1] \;
      q,\EPS[2]\subq\flt,\QFresh \;
      \G\ts\ty{Q}\ \EPS[1] <: \ty[q]{T}\ \EPS[2]
    }{
      \G[\flt]\ts t : \ty[q]{T}\ \EPS[2]
    }
  \end{minipage}\vspace{2pt}
  \begin{minipage}[t]{0.25\linewidth}\small\vspace{0pt}
    \typerule{t-ref}{
      \G[\flt]\ts t : \ty[q]{T}\ \EPS \quad \QFresh\notin q
    }{
      \G[\flt]\ts \tref~t : \ty[\QFresh]{(\TRef~\ty[q]{T})}\ \EPS
    }
  \end{minipage}%
  \begin{minipage}[t]{.32\linewidth}\small\vspace{0pt}
    \typerule{t-deref}{
      \G[\flt]\ts t : \mr{p}{T}{q}\ \EPS \; q\subq\flt
    }{
      \G[\flt]\ts !t : \ty[q]{T}\quad \EPS\EFFSEQ\JUSTU{p}
    }
  \end{minipage}%
  \begin{minipage}[t]{.43\linewidth}\small\vspace{0pt}
    \typerule{t-assgn}{
      \G[\flt]\ts t_1 : \mr{p}{T}{q}\ \EPS[1] \; \G[\flt]\ts t_2 : \ty[q]{T}\ \EPS[2]
    }{
      \G[\flt]\ts t_1 \coloneqq t_2 : \ty[\qbot]{\TUnit}\; \EPS[1]\EFFSEQ\EPS[2]\EFFSEQ\JUSTU{p}
    }
  \end{minipage}\vspace{0pt}
  \begin{minipage}[t]{.45\linewidth}\small\vspace{0pt}
    \typerule{t-abs}{
      \cx[q,x,f]{(\G\ ,\ f: \ty{F}\ ,\ x: \ty[p]{T})} \ts t : \ty{Q}\ \EPS\\
      \ty{F} = \ty[q]{\left(f(x: \ty[p]{T}) \to^{\EPS} \ty{Q}\right)}\\
      p \subq q,\QFresh \quad q \subq \flt \quad \QFresh\notin q
    }{
      \G[\flt] \ts \lambda f(x).t : \ty{F}\; \PURE
    }
  \end{minipage}%
  \begin{minipage}[t]{.52\linewidth}\small\vspace{0pt}
    \typerule{t-tabs}{
      \cx[q,x,f]{\left(\G\ ,\ f: \ty{F}\ ,\ \ty[x]{X} <: \ty[p]{T}\right)} \ts t : \ty{Q}\ \EPS \\
      F = \ty[q]{\left(\TAll{X}{x}{T}{p}{Q}{}{\EPS}\right)}\\
      p \subq q,\QFresh \quad q \subq \flt \quad \QFresh\notin q
    }{
      \G[\flt] \ts \TLam{X}{x}{T}{p}{t} : F\; \PURE
    }
  \end{minipage}\vspace{2pt}
  \begin{minipage}[t]{.43\linewidth}\small\vspace{0pt}
    \typerule{t-app}{
        \G[\flt]\ts t_1 : \ty[q]{\left(f(x: \ty[p]{T}) \to^{\EPS[3]} \ty{\ty[r]{U}}\right)}\ \EPS[1] \\
        \G[\flt]\ts t_2 : \ty[p]{T}\ \EPS[2]\quad \QFresh\notin p \\
        \ext{\QFresh \in q \Rightarrow f\notin\FX{\KILLCOMP{\EPS[3]}} \qglb r} \quad r\subq\QFresh,\flt,x,f \\
    }{
        \G[\flt]\ts t_1~t_2 : (\ty{\ty[r]{U}}\ \EPS[1]\EFFSEQ\EPS[2]\EFFSEQ\EPS[3]) [p/x, q/f]
    }
  \end{minipage}%
  \begin{minipage}[t]{.57\linewidth}\small\vspace{0pt}
    \typerule{t-app-fresh}{
      \G[\flt]\ts t_1 : \ty[q]{\left(f(x: \ty[p\,{\overlap}\, q]{T}) \to^{\EPS[3]} \ty[r]{U}\right)}\ \EPS[1] \\
      \G[\flt]\ts t_2 : \ty[p]{T}\ \EPS[2] \quad
      \QFresh \in p \Rightarrow x\notin (\FV(U),\ext{\FX{\KILLCOMP{\EPS[3]}} \qglb r}) \\
      \QFresh \in q \Rightarrow f\notin (\FV(U),\ext{\FX{\KILLCOMP{\EPS[3]}} \qglb r}) \quad r\subq\QFresh,\flt,x,f \\
    }{
      \G[\flt]\ts t_1~t_2 : ((\ty[r]{U})\ \EPS[1]\EFFSEQ\EPS[2]\EFFSEQ\EPS[3]) [p/x, q/f]
    }
  \end{minipage}
  \begin{minipage}[t]{.48\linewidth}\small\vspace{0pt}
    \typerule{t-tapp}{
      \G[\flt] \ts t : \ty[q]{\left(\TAll{X}{x}{T}{p}{Q}{}{\EPS[3]}\right)}\ \EPS[1] \\
      \QFresh \notin p \quad  f\notin\FV(U) \\
      p \subseteq \flt \quad r\subq\QFresh,\flt,x,f \quad Q = \ty[r]{U}
    }{
      \G[\flt] \ts t [ \ty[p]{T} ] : (Q\ \EPS[1]\EFFSEQ\EPS[3]) [\ty[p]T/\ty[x]{X}, q/f]
    }
  \end{minipage}%
  \begin{minipage}[t]{.52\linewidth}\small\vspace{0pt}
    \typerule{t-tapp-fresh}{
      \G[\flt] \ts t : \ty[q]{\left(\TAll{X}{x}{T}{p \overlap q}{Q}{}{\EPS[3]}\right)}\ \EPS[1] \\
      \QFresh \in p \Rightarrow x\notin\FV(U) \quad f\notin\FV(U) \\
      p \subseteq \flt \quad r\subq\QFresh,\flt,x,f \quad Q = \ty[r]{U}
    }{
      \G[\flt] \ts t [ \ty[p]{T} ] : (Q\ \EPS[1]\EFFSEQ\EPS[3]) [\ty[p]{T}/\ty[x]{X}, q/f]
    }
  \end{minipage}
  \begin{minipage}[t]{.45\linewidth}\small\vspace{0pt}
    \typerule{t-free}{
      \G[\flt]\ts t : \mr{p}{T}{q}\ \EPS
    }{
      \G[\flt]\ts \ext*{\tfree~t} : \ty[\qbot]{\TUnit}\ \EPS\EFFSEQ\JUSTK{p}
    }
  \end{minipage}%
  \begin{minipage}[t]{.55\linewidth}\small\vspace{0pt}
    \typerule{t-move}{
      \G[\flt]\ts t : \mr{p}{T}{q} \ \EPS
    }{
      \G[\flt]\ts \ext*{\tmove~t} : \mr{\QFresh}{T}{q}\ \EPS\EFFSEQ\JUSTM{p}
    }
  \end{minipage}

\caption{Typing rules of \maybelang. We omit cyclic reference types and
dual-component reference types from \Citet{Proc.ACMProgram.Lang./DengSSY25} for
simplicity. \ext*{\text{Effect-related constructs}} and \ext{\text{typing
constraints}} are highlighted.}\label{fig:maybe-typing}

\end{figure}
 
Term typing in \maybelang builds on \citet{DBLP:journals/pacmpl/WeiBJBR24,
Proc.ACMProgram.Lang./DengSSY25}. We write the judgment as \(\G[\flt] \ts t : \ty[q]{T}\
\EPS\) (see \Cref{fig:maybe-typing}). Under observation filter \(\flt\), the
term \(t\) evaluates to a value of type \(T\) with reachability qualifier \(q\)
and the evaluation may incur effect \(\EPS\). The qualifier \(q\) constrains
which parts of the environment \(t\) may reach. The effect component \(\EPS\)
tracks side effects only on entities visible through \(\flt\) and composes
sequentially in subsequent derivations. It is also one-step by default, and its
transitive closure through the context is only computed upon effect composition
(see \Cref{fig:effect-ops}). Consequently, the judgment is precise about both
the reachable results (via \(q\)) and the incurred effects (via \(\EPS\)).

\Cref{fig:maybe-typing} presents a representative subset of typing rules that
incorporate effects into the polymorphic setting. We now describe the typing
rules with a particular focus on how reachability types integrate with effects.

\subsubsection{Variables and Constants}\label{sec:formal:variables}
\Cref{typing:t-var} is pure: reading a variable produces no effects and the
result qualifier precisely names the accessed path. \Cref{typing:t-cst} is also
pure; a constant is untracked and cannot reach any location on the heap
(qualifier \(\qbot\)). 

\subsubsection{Reference Introduction}\label{sec:formal:ref-intro}
\Cref{typing:t-ref} allocates a reference but does not add effects beyond those
already incurred when evaluating the argument; allocation is tracked only by the
result qualifier, which becomes \(\QFresh\). As in
\citet{DBLP:journals/pacmpl/WeiBJBR24}, we do not allow the referent qualifier
to be fresh.

\subsubsection{Use Operations on References}\label{sec:formal:ref-use}
\Cref{typing:t-assgn} is the standard reference assignment rule, which induces a
use effect on the written reference. Intuitively, the effects observed are a
result of first evaluating the reference and then the assigned value, composed
as \(\EPS[1] \ \EFFSEQ\ \EPS[2]\), finally followed by a single write on the
reference qualifier \(p\).  The assignment rule does not induce a kill effect,
as assignment does not disable the use of the underlying reference. 
\Cref{typing:t-deref} reveals the referent's qualifier and registers exactly one
use effect on the dereferenced reference. 

\subsubsection{Reference Deallocation and Move}\label{sec:formal:ref-kill}
\Cref{typing:t-free} and \Cref{typing:t-move} are the only rules that induce a destructive 
\textit{kill} effect. 

Both first realize any prior effects of their operand, then induce the
destructive \textit{kill} effect. Note that \tmove\ first induces a \textit{use}
effect on the reference being moved. This design choice rules out illegal
``move-after-move'' and ``move-after-kill'' sequences when the effect sequencing
invariant is enforced (see \Cref{fig:effect-ops}), without needing to introduce
a separate ``move'' effect.

Another difference between the two rules is in their return types:
\Cref{typing:t-free} returns unit type, while \Cref{typing:t-move} returns the
same payload type \(T\), but with a fresh qualifier \(\QFresh\), exposing the
old reference with a unique access via a freshly allocated reference (see
\Cref{sec:formal-dynamic}).

We present unique references with atomic swap in \Cref{sec:extension}.

\subsubsection{Abstractions}\label{sec:formal:abstractions} \Cref{typing:t-abs}
is the introduction rule for abstractions: creating the abstraction does not
touch the store; any heap effects are those declared as the \textit{latent
effect} of the abstraction and occur only when the function is applied.
Following our mechanization, the rule makes two side conditions explicit. The
function's own qualifier must be non-fresh and observable in the current filter
(\(\QFresh\notin q\), \(q\subq\flt\)), and the parameter's qualifier is bounded
by the function's qualifier plus freshness (\(p \subq q,\QFresh\)): a function
may only be applied to arguments it can already reach, or to contextually fresh
ones. \Cref{typing:t-tabs} is likewise pure for type abstraction and imposes
the same bounds on the type variable's qualifier; the polymorphic body may
later perform effects when used, but none are incurred at the point of
abstraction.

\subsubsection{Application Rules}\label{sec:formal-app} \Cref{typing:t-app}
sequentializes effects in left-to-right evaluation order: first the callee
\(\EPS[1]\), then the argument \(\EPS[2]\), then the latent effects declared on
the abstraction \(\EPS[3]\). If the latent effect refers to the function self
reference \(f\) or the parameter \(x\), they are substituted with the actual
argument qualifiers \(q\) and the self reference \(p\) respectively.

\Cref{typing:t-app-fresh} is similar to \Cref{typing:t-app}, but allows the
argument qualifier to contain \(\QFresh\). It adds an additional condition that
\textit{a fresh argument must not be both returned and killed by the
abstraction}. This rules out the possibility of any ``dangling pointer''
returned from the function application that becomes untracked by our system.  

For this reason, it is also safe to regard effect qualifiers as non-fresh. Since
fresh resources are untracked, and cannot be observed by the subsequent program
unless they are returned from a function application. Therefore, as long as we
control the returning and killing of fresh resources at function boundaries, we
can safely ignore the freshness component in effect qualifiers without losing
soundness.

Both application rules enforce a similar constraint when the function captures
fresh resources (\(\QFresh \in p\)): the function body must not \textit{both
return and kill} the function self reference.

Type applications are similar, except that they do not evaluate an argument
term, and thus do not have the corresponding argument effect \(\EPS[2]\).

\subsubsection{Subsumption}\label{sec:formal:subtyping}
\Cref{typing:t-sub} permits upcasting terms to a supertype and with a coarser
set of effects, as long as the qualifier and the effects are observable in the
current context (that is, \(q,\EPS[2] \subq \varphi,\QFresh\)).

\subsection{Subtyping} \label{sec:formal-subtyping}
\begin{figure}[t]\footnotesize

\judgement{Subtyping}{\BOX{\strut\G \ts q <: q}\ \BOX{\strut \G\ts\ty{T} <: \ty{T}}\ \BOX{\strut \G\ts\ty{Q} <: \ty{Q}}\ \BOX{\strut \G\ts\ty{Q}\ \EPS <: \ty{Q}\ \EPS}}\\[1ex]
\begin{minipage}[t]{.37\linewidth}\small\vspace{0pt}
 \typerule{s-base}{\ \\}{
    \G\ts\ty{B} <: \ty{B}
  }
  \vgap
  \typerule{s-trans}{
    \G\ts\ty{T} <: \ty{S} \quad
    \G\ts\ty{S} <: \ty{U}
  }{
    \G\ts\ty{T} <: \ty{U}
  }
\end{minipage}%
\begin{minipage}[t]{.03\linewidth}
\hspace{1pt}
\end{minipage}%
\begin{minipage}[t]{.6\linewidth}\small\vspace{0pt}
  \typerule{s-ref}{
    \G\ts\ty{Q} <: \ty{R}  \quad \G\ts\ty{R} <: \ty{Q}
  }{
    \G\ts\ty{\TRef~Q} <: \ty{\TRef~R}
  }
  \vgap
  \typerule{s-fun}{
    \G\ts\ty{P} <: \ty{O} \quad
    \G\, ,\, f : \ty[\QFresh]{(f(x : O)\to^{\EPS[1]} Q)}\, ,\, x : \ty{P}\ts \ty{Q}\ \EPS[1] <: \ty{R}\ \EPS[2]
  }{
    \G\ts\ty{f(x: \ty{O}) \to^{\EPS[1]} \ty{Q}} <: \ty{f(x: \ty{P}) \to^{\EPS[2]} \ty{R}}
  }
\end{minipage}
\begin{minipage}[t]{0.3\linewidth}\small\vspace{0pt}
  \typerule{q-sub}{
    p\subq q\subq \QFresh,\dom(\G)
  }{
    \G\ts p <: q
  }
\vgap
  \typerule{q-cong}{\G\ts q_1 <: q_2}{
    \G\ts p, q_1 <: p, q_2
  }
\end{minipage}
\begin{minipage}[t]{0.3\linewidth}\small\vspace{0pt}
  \typerule{q-self}{
    f : \ty[q]{T}\in\G \quad \QFresh\notin q
  }{
    \G\ts q,f <: f
  }
  \vgap
  \typerule{q-var}{
    x : \ty[q]{T}\in\G \quad \QFresh\notin q
  }{
    \G\ts x <: q
  }
\end{minipage}
\begin{minipage}[t]{0.38\linewidth}\small\vspace{0pt}
    \typerule{q-qvar}{
      \ty[x]{X} <: \ty[q]{T} \in \Gamma \quad
      \QFresh \notin q
    }{
      \Gamma \ts p,x <: p,q
    }
    \vgap
    \typerule{q-trans}{
      \G\ts p <: q \quad
      \G\ts q <: r
    }{
      \G\ts p <: r
    }
\end{minipage}
\begin{minipage}[t]{0.35\linewidth}\small\vspace{0pt}
    \typerule{s-tvar}{
      \ty[x]{X} <: \ty[q]{T} \in \Gamma
    }{
      \Gamma \ts X <: T
    }
\end{minipage}
\begin{minipage}[t]{0.2\linewidth}\small\vspace{0pt}
    \typerule{s-top}{
      \
    }{
      \Gamma \ts T <: \TTop
    }
\end{minipage}
\begin{minipage}[t]{0.4\linewidth}\small\vspace{0pt}
  \typerule{sq-sub}{
    \G\ts\ty{S} <: \ty{T}\quad\quad \G\ts p <: q
  }{
    \G\ts\ty[p]{S} <: \ty[q]{T}
  }
\end{minipage}

\begin{minipage}[t]{0.5\linewidth}\small\vspace{0pt}
\typerule{e-sub}{
  \G\ts\FX{\alpha_1} <: \FX{\alpha_2}\quad
  \G\ts\FX{\beta_1} <: \FX{\beta_2}\quad
}{
  \G\ts\phantom{<:\,}\UKX{\alpha_1}{\beta_1} <: \UKX{\alpha_2}{\beta_2}
}
\end{minipage}%
\begin{minipage}[t]{0.5\linewidth}\small\vspace{0pt}
\typerule{sqe-sub}{
  \G\ts\ty[p]{S} <: \ty[q]{T} \qquad \G\ts \EPS[1] <: \EPS[2]
}{
  \G\ts\ty[p]{S}\ \EPS[1] <: \ty[q]{T}\ \EPS[2]
}
\end{minipage}
\caption{Subtyping rules of \maybelang.}\label{fig:maybe-subtyping}
\vspace{-3ex}
\end{figure}

\Cref{fig:maybe-subtyping} presents the subtyping rules of \maybelang, covering
qualifiers subtyping, effect subtyping, and the core type subtyping rules.

\subsubsection{Qualifier Subtyping}\label{sec:formal:qualifiers}
\Cref{typing:q-sub} introduces qualifier subtyping by inclusion, bounded by the
fresh name and the environment's domain. \Cref{typing:q-cong} is a congruence
principle: extending the right-hand component of two qualifiers preserves
subtyping. \Cref{typing:q-self} allows upcasting any function's captured
resource to itself, which is useful when reasoning about escaping closures.
\Cref{typing:q-var} traces the reachability chain and allows variables to be
upcast to their reachable qualifiers. Both \Cref{typing:q-self} and
\Cref{typing:q-var} require the qualifier in context to be non-fresh.
\Cref{typing:q-qvar} lifts bounds on qualified type variables, and
\Cref{typing:q-trans} provides transitivity for qualifiers, composing subtyping
steps. \Cref{typing:sq-sub} combines subtyping of the underlying types with
subtyping of their qualifiers: if \(S <: T\) and \(p <: q\), then \(\ty[p]{S} <:
\ty[q]{T}\). The rule is pointwise and imposes no further interaction beyond
well-formedness of the qualifiers involved.

\subsubsection{Effect Subtyping}\label{sec:formal:effects} \Cref{typing:e-sub} defines
effect subtyping componentwise. With effects being represented as a pair (a use
component \(\alpha\) and a kill component \(\beta\)), subtyping requires each
component to be in a sub-qualifier relation. \Cref{typing:sqe-sub} is the combined subtyping judgment over type, qualifier,
and effects. Each component must advance in its own subtyping relation: the
underlying type by type subtyping, the qualifier by qualifier subtyping, and the
effects by effect subtyping.

\subsubsection{Core Type Subtyping}\label{sec:formal:core-subtyping}
\Cref{typing:s-base} states reflexivity for base types: each ground type is a
subtype of itself. \Cref{typing:s-trans} provides transitivity, allowing
multi-step derivations to collapse into a single subtyping step.

\subsubsection{Reference Types Subtyping}\label{sec:formal:ref-subtyping}
\Cref{typing:s-ref} makes reference types \emph{invariant} in their payload: it
demands both \(Q <: R\) and \(R <: Q\). This rules out unsound variance for
writable references. Intuitively, two reference types are related by subtyping
only when their qualified payloads are mutually subtypes. Subtyping does not grant
additional read/write capability.

\subsubsection{Function Types Subtyping}\label{sec:formal:fun-subtyping}
\Cref{typing:s-fun} gives the standard contravariance in the parameter and
covariance in the result, extended to our effect annotations. The premise \(P <:
O\) captures contravariant function arguments, and the arguments must be pure.
The body check is performed under a context extended with a fresh self name
\(f\) and parameter \(x:P\). Both the qualified type of the result and the
latent effect are checked covariantly under this context. 

\subsubsection{Type Variables and Top}\label{sec:formal:tvar-top}
\Cref{typing:s-tvar} reads bounds from the context: if the environment contains
\(X <: T\), then \(X\) is a subtype of \(T\). \Cref{typing:s-top} introduces the
top type: every type is a subtype of \(\TTop\).

\subsection{Dynamics} \label{sec:formal-dynamic}

\begin{figure}\footnotesize

\judgement{Syntax, Store, Store Typing, and Evaluation Contexts}{\BOX{\maybelang}}
\[\begin{array}{l@{\qquad\qquad}l@{\qquad\qquad}l@{\qquad\qquad}l}
    \ell,\ell_1,\ell_2 & \in & \Loc                                                                        & \text{Locations}           \\
    p,q,r,w            & \in & \mathcal{P}_{\mathsf{fin}}(\Var \uplus \Loc \uplus \{ \QFresh \})           & \text{Qualifiers}          \\
    \flt               & \in & \mathcal{P}_{\mathsf{fin}}(\Var \uplus \Loc)                                & \text{Observations}        \\ [.5ex]
    t                  & ::= & \cdots \mid \ell                                                            & \text{Terms}               \\
    v                  & ::= & \lambda f(x).t \mid {c} \mid {\ell} \mid \tunit \mid \Lambda f(\ty[x]{X}).t & \text{Values}              \\ [.5ex]
    v_{\DEAD}          & ::= & v \mid \DEAD                                                                & \text{Store Binding}       \\
    {\sigma}           & ::= & \varnothing \mid \sigma,{\ell:v_\DEAD}                                      & \text{Store}               \\
    \Sigma             & ::= & \varnothing \mid \Sigma,\ell : (0, \mathsf{ref}, T, q)                      & \text{Store Typing}        \\ [.5ex]
    {S}                & ::= & \sigma\, \square\, \sigma                                                   & \text{Store Contexts}      \\
    {E}                & ::= & \square \mid E\ t \mid v\ E \mid \tref~E \mid\ !{E} \mid {E} := {t}         &                            \\
                       &     & \mid {v} := {E}\mid \tfree~E\mid \tmove~E                                   & \text{Evaluation Contexts} \\
\end{array}
\]

\judgement{Term Typing, Well-Formed and Well-Typed Stores}{
    \BOX{\cx[\flt]{[\Gamma\mid\Sigma]} \ts t : \ty{Q}}\ 
    \BOX{\WF{\Sigma}}\
    \BOX{\cx[\varphi]{[\Gamma \mid \Sigma]} \mid k \vdash \sigma}
}\\\vspace{-8pt}

\[\begin{array}{lll@{\qquad\qquad}lll}
    \DOM(\Sigma) & := & \{\ell \mid \ell : (0, \mathsf{ref}, T, q) \in \Sigma\} & \DOM(\sigma) & := & \{\ell \mid \ell : v \in \sigma\} \\
\end{array}\]
\vspace{-10pt}
\begin{minipage}[t]{.4\linewidth}\vspace{-2pt}
    \typerule{t-loc}{
        \Sigma(\ell) = (0, \mathsf{ref}, T, q)\quad q\subq\DOM(\Sigma) \\
        \FV(T)=\varnothing\quad\FTV(T)=\varnothing \quad q,\ell \subq \flt
    }{
        \cx[\flt]{[\G\mid\Sigma]}\ts \ell : \ty[\ell]{\TRef~[\ty[q]{T}]}\ \PURE
    }
\end{minipage}%
\begin{minipage}[t]{.2\linewidth}\vspace{0pt}
    \typerule{st-emp}{\phantom{}\\\phantom{}}{\WF{\varnothing}}\qquad\qquad
\end{minipage}%
\begin{minipage}[t]{.4\linewidth}\vspace{0pt}
    \typerule{st-con}
    {\WF{\Sigma}\quad \FV(T)=\varnothing \quad \FTV(T)=\varnothing \\
    q \in \DOM(\Sigma) \quad \ell \notin\DOM(\Sigma) }{\WF{\Sigma\, ,\, \ell : (0, \mathsf{ref}, T, q)}}
\end{minipage}

\[\begin{array}{r@{\quad}l}
    \cx[\varphi]{[\Gamma \mid \Sigma]} \mid k \vdash \sigma := 
    & \DOM(\sigma) \subseteq \DOM(\Sigma) \land |\sigma| = |\Sigma| \land (\varphi \setminus k \subseteq \DOM(\sigma)) \land (k \subseteq \DOM(\Sigma)) \land \mbox{} \\
    & \forall \ell.\ (\ell \in k \implies \sigma(\ell) = \DEAD) \land \mbox{} \\
    & \phantom{\forall \ell.\ } \bigl(\ell \in \varphi \setminus k \land \Sigma(\ell) = (n, \mathit{flag}, T, q) \land \sigma(\ell) = v \implies \cx[\varphi]{[\Gamma \mid \Sigma]} \vdash v : \ty[q]{T}\bigr) \\
\end{array}\]

\judgement{Reduction Rules}{\BOX{\sigma \mid t \redv \sigma\mid t}}\\\vspace{-5pt}
\semrulelabel{step-beta} \semrulelabel{step-ref} \semrulelabel{step-deref} \semrulelabel{step-assign} \semrulelabel{step-free} \semrulelabel{step-move}
\[\begin{array}{r@{\,\mid\,}l@{\quad}c@{\quad}r@{\,\mid\,}l@{\qquad}r}
  \sigma                         & \CX[gray]{E}{(\lambda f(x).t)\ v} & \redv & \sigma                                      & \CX[gray]{E}{t[v/x,(\lambda f(x).t)/f]}                                & \rulename*{step-$\beta$} \\ [1ex]
  \sigma                         & \CX[gray]{E}{\tref~v}             & \redv & \sigma,\,{\ell : v}                         & \CX[gray]{E}{\ell}, \qquad \ell \not\in \DOM(\sigma)                   & \rulename*{step-ref}     \\ [1ex]
  \CX[gray]{S}{\ell : v}         & \CX[gray]{E}{!\ell}               & \redv & \CX[gray]{S}{\ell : v\phantom{'}}           & \CX[gray]{E}{v}                                                        & \rulename*{step-deref}   \\ [1ex]
  \CX[gray]{S}{\ell : v}         & \CX[gray]{E}{\ell := v'}          & \redv & \CX[gray]{S}{\ell : v'}                     & \CX[gray]{E}{\tunit}                                                   & \rulename*{step-assign}  \\ [1ex]
  \CX[gray]{S}{\ell : v_{\DEAD}} & \CX[gray]{E}{\tfree~\ell}         & \redv & \ext{\CX[gray]{S}{\ell : \DEAD\phantom{'}}} & \CX[gray]{E}{\tunit}                                                   & \rulename*{step-free}    \\ [1ex]
  \CX[gray]{S}{\ell : v}         & \CX[gray]{E}{\tmove~\ell}         & \redv & \ext{\CX[gray]{S}{\ell : \DEAD}\,\ell_1:v}  & \ext{\CX[gray]{E}{\ell_1}}, \qquad \!\!\!\!\ell_1 \not\in \DOM(\sigma) & \rulename*{step-move}    \\ [1ex]
  \end{array}\]
  \vspace{-8pt}

  \judgement{Multi-Step Reduction}{\BOX{\sigma \mid t \redv^{n} \sigma' \mid t'}}\\\vspace{-1pt}
  \begin{minipage}[t]{.4\linewidth}
      \typerule{step-0}
      {\strut}
      {\sigma \mid t \redv^{0} \sigma \mid t}
  \end{minipage}
  \begin{minipage}[t]{.55\linewidth}
      \typerule{step-succ}
      {\sigma \mid t \redv \sigma_1 \mid t_1 \quad
      \sigma_1 \mid t_1 \redv^{n} \sigma' \mid t'}
      {\sigma \mid t \redv^{n+1} \sigma' \mid t'}
  \end{minipage}

  \caption{Store Typing, Well-Formed Stores, and Operational Semantics for
      \maybelang. We highlight the semantics for \ext{\tfree \text{ and }\tmove}.
  }\label{fig:maybe-semantics}

  \vspace{-3ex}
\end{figure}
 
The dynamic semantics of \maybelang\ follows the standard call-by-value
reduction for the \(\lambda\)-calculus with mutable references.  A runtime
configuration pairs a store \(\sigma\) with an expression, and reduction steps
are taken in the evaluation contexts from \Cref{fig:maybe-semantics}.  Stores
map locations to either a value or the distinguished marker \(\DEAD\) denoting
deallocated states, which is essential for soundness with qualified types. The
store typing judgments in the figure ensure that every allocated location is
described by a well-formed qualified reference type and that any location
visible to the observer set \(\flt\) is inhabited by a value consistent with its
typing annotation.

The core reduction rules are those of the traditional CBV calculus. Rule
\Cref{rule:step-beta} performs term-level function application by substituting
the argument value for the parameter and the closure for its self-reference.
Reference allocation \Cref{rule:step-ref} extends the store with a fresh
location whose content is the allocated value (see \Cref{fig:store-ref}); \Cref{rule:step-deref} reads the
value stored at a reachable location; and \Cref{rule:step-assign} overwrites the
content of an existing location while preserving the overall store domain.
Because evaluation contexts enforce left-to-right evaluation, these operations
align with the sequencing discipline required by the static effect system.

The highlighted rules \Cref{rule:step-free} and \Cref{rule:step-move} capture
the destructive effects that distinguish \maybelang. Deallocation
(\(\tfree~\ell\)) replaces the content of the reference cell corresponding to
\(\ell\) with \(\DEAD\) (see \Cref{fig:store-free}). Note that it does that
regardless of whether the cell is already deallocated or not so that the
deallocation operation is \textit{idempotent} (see
\Cref{sec:motivation-effects-idempotent}). Move (\(\tmove~\ell\)), on the other
hand, is not idempotent, and requires a live location, then allocates a fresh
location \(w\) that stores the payload while marking the original cell as
\(\DEAD\) (see \Cref{fig:store-move}). The term-level result is the fresh
location, which matches the typing rule that returns the moved value at a fresh
qualifier.  The freshness side condition in \Cref{rule:step-move} enforces the
same dynamic discipline as the static typing rule: a moved reference cannot be
accessed again through its original pointer, yet the payload remains available
through a newly allocated name. 

We describe the operational semantics of unique references (\Cref{rule:step-uref})
and atomic swaps (\Cref{rule:step-swap}) in \Cref{sec:extension}.
\Cref{fig:maybe-store-evolution} in \Cref{app:store-evolution} illustrates these
heap operations step by step on concrete store cells.

\begin{figure}[t]
\footnotesize

\judgement{Qualifier Transitive Lookup and Overlap}{\BOX{{\color{gray}\G\vdash}\, \qtrans[n]{q}}\ \BOX{{\color{gray}\G\vdash}\, \qtrans{q}}}\ \BOX{{\color{gray}\G\vdash}\, p \overlap q}
\[\begin{array}{l@{\ \,}c@{\ \,}l@{\qquad\qquad\qquad\ \ }l}
    {\color{gray}\G\vdash}\, \qtrans[0]{q}   & := & q                                                                                   & \text{Qualifier Transitive Lookup Base Case}      \\ [1.1ex]
    {\color{gray}\G\vdash}\, \qtrans[n+1]{q} & := & \qtrans[n]{((\bigcup_{x \in q}\, \left\{\, y \mid x \reaches y\, \right\}) \cup q)} & \text{Qualifier Transitive Lookup Inductive Case} \\ [1.1ex]
    {\color{gray}\G\vdash}\, \qtrans{q}      & := & \qtrans[\norm{\G}]{q}                                                               & \text{Qualifier Full Transitive Lookup}           \\ [1.1ex]
  {\color{gray}\G\vdash}\, p \overlap q      & := & \QFresh, {(\qtrans{p} \qglb\, \qtrans{q})}                                          & \text{Qualifier Overlap}                          \\ [1.1ex]
\end{array}\]

\judgement{Effect Composition}{\BOX{{\color{gray}\G\vdash}\, \EPS[1]\EFFSEQ\EPS[2]}\ \BOX{\EPS[1]\EFFJOIN\EPS[2]}}\

\begin{minipage}[t]{.52\linewidth}\vspace{0pt}
\typerule{eff-seq}{
  {\color{gray}\G\vdash}\, \FX{\KILLCOMP{\EPS[1]} \overlap \USECOMP{\EPS[2]}\subq \QFresh}\qquad
}{
  {\color{gray}\G\vdash}\, \EPS[1]\EFFSEQ\EPS[2] = \UKX
    {(\USECOMP{\EPS[1]},\USECOMP{\EPS[2]})}
    {(\KILLCOMP{\EPS[1]},\KILLCOMP{\EPS[2]})}
}
\end{minipage}%
\begin{minipage}[t]{.46\linewidth}\vspace{0pt}
\typerule{eff-join}{\phantom{}}{
  \EPS[1]\EFFJOIN\EPS[2] = \UKX
    {(\USECOMP{\EPS[1]},\USECOMP{\EPS[2]})}
    {(\KILLCOMP{\EPS[1]},\KILLCOMP{\EPS[2]})}
}
\end{minipage}

\caption{Operations on effects. Sequential effect composition (\EFFSEQ)
is the component-wise set union and is only defined if no
variable/location is used after being killed. The non-sequential effect
join (\EFFJOIN) always computes the same union without requiring disjoint use
and kill sets.}\label{fig:effect-ops}

\vspace{-3ex}
\end{figure}

\subsection{Metatheory} \label{sec:formal-meta}

The metatheory distills how the operational semantics respects the
discipline imposed by effects.  We first summarize how single reductions
grow the store while maintaining typing, then extend the reasoning to whole
evaluation traces, and finally argue that disjoint threads remain safe when
they execute in parallel. We state the main results here and defer detailed
intuitions and proof sketches to \Cref{app:proofs}; all results are mechanized in
Rocq.

\paragraph{Single-step execution.} The progress theorem rules out stuck states:
a closed, well-typed term is either already a value or can take a reduction step
in any well-typed store compatible with its remaining effect budget.
\needspace{8\baselineskip}
\begin{theorem}[Progress]\label{thm:progress}\ \vspace{-5pt}
  \infrule{
    \csx[\DOM(\Sigma)]{\varnothing}{\Sigma}  \ts t : \ty[q]{T}\ \EPS \qquad
    \WF{\Sigma} \qquad
    \csx[\DOM(\Sigma)]{\varnothing}{\Sigma}\mid k \ts \sigma \qquad
    {\color{gray}\G\vdash}\, \JUSTK{k}\EFFSEQ\EPS
  }{
    t\text{ value}
    \quad\lor\quad
    \exists\,\sigma',t'.\ \sigma \mid t \redv \sigma' \mid t'
  }
\end{theorem}

\paragraph{Type preservation.}
The preservation theorem captures the obligations discharged by a single
reduction step, including how the store grows, which locations may be killed,
and how residual effects remain compatible with future reasoning.
\begin{theorem}[Preservation]\label{thm:preservation}\ \vspace{-8pt}
  \infrule{
    \csx[\DOM(\Sigma)]{\varnothing}{\Sigma}  \ts t : \ty[q]{T}\ \EPS \qquad
    \sigma\mid t \redv \sigma' \mid t' \\ [1ex] 
    \WF{\Sigma} \qquad 
    \csx[\DOM(\Sigma)]{\varnothing}{\Sigma}\mid k \ts \sigma \qquad
    {\color{gray}\G\vdash}\, \JUSTK{k}\EFFSEQ\EPS
  }{
		\DOM(\sigma) \subq \DOM(\sigma') \quad 
		\exists\,\Sigma',\EPSPR,\FX{\EPS''}\,p,\,\,k',\,\varphi'.\quad
    \FX{\EPS''} \subq \EPS \quad
		\Sigma' \supseteq \Sigma \quad 
		\varphi \subq \varphi' \quad 
		\WF{\Sigma'} \\[1px]
		k' \subq \FX{\KILLCOMP{\EPS}} \quad 
		\csx[\DOM(\Sigma')]{\varnothing}{\Sigma'} \mid k,k'\ts \sigma' \quad 
		p,\EPSPR\subq\DOM(\Sigma') \setminus \DOM(\Sigma) \\[1ex]
		\FX{\KILLCOMP{\EPSPR}} \qglb p \subq \qbot \quad
		{\color{gray}\G\vdash}\, \JUSTK{k,k'}\EFFSEQ\EPS,\EPSPR \quad
		\csx[\DOM(\Sigma')]{\varnothing}{\Sigma'} \ts t' : \ty[{q [p/\QFresh]}]{T}\ \FX{\EPS''\EFFSEQ \EPSPR} \\[1ex]
  }
\end{theorem}

\paragraph{Multi-step execution.} Building on preservation and progress,
the next lemma shows that every closed, well-typed term either keeps stepping
or eventually produces a value when assuming classical reasoning.
\begin{lemma}[Multi-step Type Safety]\label{lem:multi-step-type-safety}\
  \infrule{
    \csx[\varphi]{\varnothing}{\Sigma}  \ts t : \ty[q]{T}\ \EPS \qquad
    \forall P.~ P \lor \neg P
  }{
    \left(
      \forall n.~\exists t',\sigma'.~ t \mid \varnothing \redv^{n} t' \mid \sigma'
    \right)
    \quad\lor\quad
    \left(
      \exists n,\sigma',v.~ t \mid \sigma \redv^{n} v \mid \sigma' \land v \text{ value}
    \right)
  }
\end{lemma}

\paragraph{Parallel composition.} Preservation extends compositionally to two threads
that observe disjoint portions of the store. When their static effect budgets
are compatible with the available compensation, each thread can reduce on its
restricted heap slice while maintaining a well-typed store and without
overlapping the other thread's residual uses or kills.

\vspace{1pt}
\begin{corollary}[Parallel Progress and Preservation with Effects]\label{cor:parallel-progress}\ \vspace{-8pt}
  \infrule{
    \begin{array}{l@{\quad}l@{\quad}l}
      \cx[\varphi_1]{[\varnothing \mid \Sigma]} \ts t_1 : \ty[q_1]{T_1}\ \EPS[1]
        & \cx[\varphi_1]{[\varnothing \mid \Sigma]} \mid k_1 \ts \sigma
        & t_1, t_2\text{ non-value} \quad
          \varphi_1 \overlap \varphi_2 \subq \{\vardiamondsuit\} \quad
          \WF{\Sigma} \\[1ex]
      \cx[\varphi_2]{[\varnothing \mid \Sigma]} \ts t_2 : \ty[q_2]{T_2}\ \EPS[2]
        & \cx[\varphi_2]{[\varnothing \mid \Sigma]} \mid k_2 \ts \sigma
        & {\color{gray}\G\vdash}\, \JUSTK{k_1}\EFFSEQ\EPS[1] \quad
        {\color{gray}\G\vdash}\, \JUSTK{k_2}\EFFSEQ\EPS[2]
    \end{array}
  }{
    \exists\, t_1'\,t_2'\,\sigma_1'\,\sigma_2'\,\Sigma_1\,\Sigma_2\,\varphi_1'\,\varphi_2'\,\EPSPR[1]\,\EPSPR[2]\,k_1'\,k_2'\,p_1\,p_2.\quad 
        p_1 \overlap p_2 \subq \{\vardiamondsuit\}
    \\[1ex]
    \begin{array}{l@{\quad}l@{\quad}l@{\quad}l}
      t_1 \mid \sigma_{\restriction_{\varphi_1}} \redv t_1' \mid \sigma_1'
        & \cx[\varphi_1']{[\varnothing \mid \Sigma_1]} \mid k_1' \ts \sigma_1'
        & \cx[\varphi_1']{[\varnothing \mid \Sigma_1]} \ts t_1' : \ty[p_1]{T_1}\ \EPSPR[1]
        & \Sigma_1 \supseteq \Sigma \\[1ex]
      t_2 \mid \sigma_{\restriction_{\varphi_2}} \redv t_2' \mid \sigma_2'
        & \cx[\varphi_2']{[\varnothing \mid \Sigma_2]} \mid k_2' \ts \sigma_2'
        & \cx[\varphi_2']{[\varnothing \mid \Sigma_2]} \ts t_2' : \ty[p_2]{T_2}\ \EPSPR[2]
        & \Sigma_2 \supseteq \Sigma \\[1ex]
    \end{array}
  }
\end{corollary}

\subsection{Semantic Type Soundness} 

In addition to proving syntactic type soundness, we define unary logical
relations for a variant of the $\maybelang{}$-calculus with kill effect, and
prove semantic type soundness (the fundamental property). For simplicity, the
variant excludes subtyping, following prior modeling of those features in other
settings~\cite{bao_modeling_2025}. Interested readers can find our Rocq
mechanization online, and we provide the complete development as supplemental
material accompanying this paper.

\section{Extension: Unique References and Atomic Swap}\label{sec:extension}

The base calculus already supports familiar reference idioms. A \(\tget\)
operation (\eg, \Cref{typing:t-deref}) leaves the store unchanged, while a
\(\tput\) operation (\eg, \Cref{typing:t-assgn}) performs a write. We extend
these ideas to ``unique'' references whose current payload can be extracted only
through take-and-replace operations.
We introduce a \(\tswap\) operation (\eg, \Cref{typing:t-swap} in \Cref{fig:maybe-swap}) that atomically returns a fresh inner
referent, while ``moving'' a new payload into the reference.  This extension
mirrors systems-language constructs, while preserving the effect discipline from
\Cref{sec:formal-meta}.

\begin{figure}[t]\footnotesize
  \judgement{Syntax}{\BOX{{\mathsf{F}_{\varepsilon <:\exists! }^{\vardiamondsuit}}}}\
\[\begin{array}{l@{\qquad}l@{\qquad}l@{\qquad}l}
    t,t_1,t_2   & ::= & \cdots \mid \turef~t \mid \tswap~t_1~t_2                     & \text{Terms}               \\
    \Sigma      & ::= & \varnothing \mid \Sigma,\ell : (\ext{n}, \mathsf{ref}, T, q) & \text{Store Typing}        \\
    {E}         & ::= & \cdots \mid \turef~E\mid \tswap~E~t\mid \tswap~t~E           & \text{Evaluation Contexts} \\
\end{array}\]
  \judgement{Unique Reference Notation}{\BOX{\TRefpow{n}{\ty[q]{T}}}}\
  \[
  \ext*{\TRefpow{0}{\ty[q]{T}}} := \ty[q]{T}
  \qquad\qquad
  \ext*{\TRefpow{n+1}{\ty[q]{T}}} := \TRef~\ty[\QFresh]{(\TRefpow{n}{\ty[q]{T}})}
  \]
  \judgement{Term Typing (with Unique References)}{\BOX{\strut\G[\flt] \ts t : \ty[q]{T}\ \EPS}}\\[1ex]
  \begin{minipage}[t]{.5\linewidth}\vspace{0pt}
  \typerule{t-uloc}{
    \Sigma(\ell) = (\ext{n}, \mathsf{ref}, T, q)\quad \FV(T)=\varnothing\quad\FTV(T)=\varnothing \quad \ell \subq \flt
  }{
    \cx[\flt]{[\G\mid\Sigma]}\ts \ell : \ext*{\ty[\ell]{\TRefpow{n}{(\TRef~\ty[q]{T})}}}\ \PURE
  }
  \end{minipage}%
  \begin{minipage}[t]{.45\linewidth}\vspace{0pt}
    \typerule{t-uref}{ 
      \G[\flt]\ts t : \ty[r]{(\TRef~\ty[q]{T})}\ \EPS
    }{
      \G[\flt]\ts \turef~t : \ty[\QFresh]{(\TRef~\ty[\QFresh]{(\TRef~\ty[q]{T})})}\ \EPS\EFFSEQ\JUSTM{r}
    }
  \end{minipage}
  \begin{minipage}[t]{\linewidth}\vspace{0pt}
    \typerule{t-swap}{ 
      \G[\flt]\ts t_1 : \ty[p]{(\TRef~\ty[\QFresh]{(\TRef~\ty[q]{T})})}\ \EPS[1] \quad
      \G[\flt]\ts t_2 : \ty[r]{(\TRef~\ty[q]{T})}\ \EPS[2]
    }{ 
      \G[\flt]\ts \tswap~t_1~t_2: \ty[\QFresh]{(\TRef~\ty[q]{T})}\quad \EPS[1]\EFFSEQ\EPS[2]\EFFSEQ\UKX{p,r}{r}
    }
  \end{minipage}

\judgement{Reduction Rules}{\BOX{\sigma \mid t \redv \sigma\mid t}}
\semrulelabel{step-uref} \semrulelabel{step-swap}
\[\begin{array}{r@{\,\mid\,}l@{\quad}c@{\quad}r@{\,\mid\,}l@{\qquad}r}
  \CX[gray]{S}{\ell_1 : v}                             & \CX[gray]{E}{\turef~\ell_1}          & \redv & \CX[gray]{S}{\ell_1 : \DEAD}\,\ell:v                                 & \CX[gray]{E}{\ell}, \qquad \!\!\!\!\ell \not\in \DOM(\sigma) & \rulename*{step-uref} \\ [1ex]
  \CX[gray]{S}{\ell_1 : v_1}\CX[gray]{S}{\ell_2 : v_2} & \CX[gray]{E}{\tswap~\ell_1~\ell_2}   & \redv & \CX[gray]{S}{\ell_1 : v_2}\CX[gray]{S}{\ell_2 : \DEAD}\,\ell:v_1     & \CX[gray]{E}{\ell}, \qquad \!\!\!\!\ell \not\in \DOM(\sigma) & \rulename*{step-swap} \\ [1ex]
  \end{array}\]\vspace{-8pt}

\caption{Typing rules for unique references with fresh inner qualifiers and
atomic swap. We encode \ext*{\text{nested unique references}} with a \ext{\text{depth parameter}} to expose
the terminal type and qualifier.}
\label{fig:maybe-swap}
\vspace{-3ex}
\end{figure}
 
\subsection{Unique References}

A unique reference is a reference whose referent qualifier is \(\QFresh\). This
alone distinguishes unique references from normal references because the
referent qualifier of the latter can never be fresh (see \Cref{typing:t-ref}).
The inner fresh qualifier guarantees that the stored payload is unobservable
from the surrounding context until it is explicitly retrieved.  

Unique references can be nested, and we define the family
\(\TRefpow{n}{\ty[q]{T}}\) (see \Cref{fig:maybe-swap}) to compactly represent
unique references of arbitrary depths. This allows us to maintain a transparent
view on the terminal type and qualifier, preserving a telescoping structure in
store typing (see \Cref{typing:st-con} in \Cref{fig:maybe-semantics}) without
circular pointers even in the presence of swap operations.

To support this telescoping structure uniformly across standard references, unique
references, and subsequently linked lists, we refine the store typing \(\Sigma\)
to map each location to a four-tuple \((n, \mathit{flag}, T, q)\). The
\(\mathit{flag}\) distinguishes whether the location holds a standard/unique
reference (\textsf{ref}) or a linked list (\textsf{list}). The integer \(n\)
tracks the nesting depth of unique ownership, where \(n=0\) corresponds to a
standard, non-unique reference. The components \(T\) and \(q\) continue to
denote the underlying payload type and reachability qualifier.

As shown in \Cref{fig:maybe-swap}, we augment the surface syntax with a
dedicated constructor \turef\ for unique references. The typing rule \Cref{typing:t-uref} moves the referent
into a newly allocated reference and records a single \textit{move} effect
\(\JUSTM{d}\).  This treatment mirrors the linear resource discipline already
used for \tmove: a unique reference behaves like an owned box whose contents may
be retrieved only by replacement, preserving a unique access path.

The store typing discipline for unique references ensures unique access to the
referent: \turef\ allocates a unique reference by
incrementing the unique-depth counter on the resulting reference, while
simultaneously deallocating the original payload, thus ensuring unique access to
the referent. The corresponding formal operational semantics are presented
with \Cref{rule:step-uref} in \Cref{fig:maybe-swap}.

\subsection{Atomic Swap}

The \tswap\ primitive builds on the same mechanism as \turef; instead of
moving a value into a new unique reference, it moves a new payload into an
existing one, consuming the incoming resource. It reinitializes the store cell
with the incoming argument and returns the previous payload as a freshly
allocated cell. This embodies a ``take-and-replace'' discipline: removing the
current payload is only permitted when we simultaneously provide a replacement,
so the unique reference never points to an uninitialized slot. No other
expression can observe the stored value until a swap occurs. The corresponding
formal operational semantics are presented with \Cref{rule:step-swap} in
\Cref{fig:maybe-swap}.

To showcase this, we consider Rust's \texttt{swap} on \textsf{Box} (see
\Cref{fig:swapbox}), which performs an ownership exchange: the caller
\textit{uses} both unique references and exchanges their underlying content, but
never \textit{kills} them, so the two references remain unaliased after the
exchange. We assume a Scala-style \texttt{Default} type class that supplies a
canonical placeholder payload via \texttt{default[T]}, and that \texttt{swapBox}
receives implicit evidence witnessing that $\texttt{T}\trackvar{q}$ is an instance
of \texttt{Default}. The function takes two unique references \texttt{r1} and
\texttt{r2}, allocates a temporary reference for the exchange, uses three swaps to
exchange the payloads, and finally deallocates the temporary reference; it
\textit{uses} both references but \textit{kills} neither, so they remain unaliased
after the exchange.

\begin{figure}[ht]
  \centering
  \begin{varwidth}{\linewidth}
\begin{lstlisting}
  def swapBox[T$\trackvar{q}$](r1: Ref[Ref[T$\trackvar{q}$]$^\QFresh$]$\trackvar{r1}$, r2: Ref[Ref[T$\trackvar{q}$]$^\QFresh$]$\trackvar{r2}$) (using Default[T$\trackvar{q}$]) : Unit$^\lstcm{\qbot}$ $\lstcm{\JUSTU{\texttt{r1,r2}}} $ = {
    val tmp = new Ref(default[T$\trackvar{q}$]) // : Ref[Ref[T$\lstcm{\trackvar{q}}$]$^\lstcm{\QFresh}$]$^\lstcm{\trackvar{tmp}}$ $\PURE$
    val tmp1 = swap(r1, tmp)       // : Ref[T$\lstcm{\trackvar{q}}$]$^\lstcm{\QFresh}$ $\UKX{\texttt{r1,tmp}}{\texttt{tmp}}$
    val tmp2 = swap(r2, tmp1)      // : Ref[T$\lstcm{\trackvar{q}}$]$^\lstcm{\QFresh}$ $\UKX{\texttt{r2,tmp1}}{\texttt{tmp1}}$
    val tmp3 = swap(r1, tmp2)      // : Ref[T$\lstcm{\trackvar{q}}$]$^\lstcm{\QFresh}$ $\UKX{\texttt{r1,tmp2}}{\texttt{tmp2}}$
    free(tmp3)                     // : Unit$^\lstcm{\qbot}$ $\JUSTK{\texttt{tmp3}}$
  }
\end{lstlisting}
  \end{varwidth}
  \caption{Rust-style \texttt{swap} on \texttt{Box} using unique references.}
  \label{fig:swapbox}
\end{figure}

\section{Extension: Unique Linked Lists}\label{sec:extension-lists}

We demonstrate a linked list implementation based on a specialized
version of unique references and swap (see \Cref{fig:maybe-list}). Algebraically, 
the unique list type can be understood as:
\[
  \TList~\ty[q]{T} \approx \TUnit + (\ty[q]{T} \times \TRef~\ty[\QFresh]{(\TList~\ty[q]{T})})
\]
In this encoding, each node is represented as a unique reference to the rest of the list, 
which is uniquely owned by its parent node. The tail of the list is thus uniquely owned 
by the head and can only be accessed by swapping it out. 

For example, a list containing elements \texttt{[1, 2, 3]} is constructed by chaining 
\texttt{cons} and \texttt{uref} (which allocates a unique reference) operations:
\begin{lstlisting}
  val l = uref(cons(1, uref(cons(2, uref(cons(3, uref(nil)))))))
\end{lstlisting}

We defer detailed examples of structural manipulation on unique linked
lists--such as recursive summation and destructive merging via destructive
pattern matching--to the case studies in \Cref{sec:casestudy}.

\begin{figure}[t]\footnotesize
  \judgement{Syntax Extensions for Lists}{\BOX{\maybelang}}\
\[\begin{array}{l@{\qquad}l@{\qquad}l@{\qquad}l}
    T           & ::= & \cdots \mid \ext{\TList~\ty[\qbot]{T}}                              & \text{Types}               \\
    t           & ::= & \cdots \mid \tnil \mid \tcons~t_1~t_2 \mid \tmatchlist{t_1}{t_2}{x~y \to t_3} & \text{Terms} \\
    v           & ::= & \cdots \mid \ext*{\tnil \mid \tcons~v~\ell}                         & \text{Values} \\
    v_L         & ::= & \ext*{[] \mid v :: v_L}                                             & \text{List Store Binding} \\
    v_{\DEAD}^+ & ::= & v \mid \ext*{v_L} \mid \DEAD                                        & \text{Store Binding}       \\
    {\sigma}    & ::= & \varnothing \mid \sigma,{\ell:v_{\DEAD}^+}                          & \text{Store}               \\
    \Sigma      & ::= & \cdots \mid \Sigma, \ell : (n, \ext*{\mathsf{list}}, T, \qbot)      & \text{Store Typing}        \\
    {E}         & ::= & \cdots \mid \tcons~E~t \mid \tcons~t~E \mid \tmatchlist{E}{t_2}{x~y \to t_3} & \text{Evaluation Contexts} \\
\end{array}\]

  \judgement{Term Typing and Subtyping (Lists)}{\BOX{\strut\G[\flt] \ts t : \ty[q]{T}\ \EPS}\ \ \BOX{\strut \G\ts\ty[\qbot]{S} <: \ty[\qbot]{T}}}\\[1ex]
  \begin{minipage}[t]{.2\linewidth}\vspace{0pt}
    \typerule{t-nil}{
      \strut \\
      \strut
    }{
      \G[\flt]\ts \tnil : \TList~\ty[\qbot]{T}\ \PURE
    }
  \end{minipage}%
  \begin{minipage}[t]{.35\linewidth}\vspace{0pt}
    \typerule{t-cons}{
      \G[\flt]\ts t_1 : \ty[\qbot]{T}\ \EPS[1] \\
      \G[\flt]\ts t_2 : \ty[d_2]{(\TRef~\ty[\QFresh]{(\TList~\ty[\qbot]{T})})}\ \EPS[2]
    }{
      \G[\flt]\ts \tcons~t_1~t_2 : \ty[d_2]{(\TList~\ty[\qbot]{T})}\ \EPS[1]\EFFSEQ\EPS[2]
    }
  \end{minipage}%
  \begin{minipage}[t]{.45\linewidth}\vspace{0pt}
    \typerule{t-swap-list}{
      \G[\flt]\ts t_1 : \ty[p]{(\TRef~\ty[\QFresh]{(\TList~\ty[\qbot]{T})})}\ \EPS[1] \\
      \G[\flt]\ts t_2 : \ty[r]{(\TList~\ty[\qbot]{T})}\ \EPS[2]
    }{
      \G[\flt]\ts \tswap~t_1~t_2 : \ty[\QFresh]{(\TList~\ty[\qbot]{T})}\\ 
      \EPS[1]\EFFSEQ\EPS[2]\EFFSEQ\UKX{p,r}{r}
    }
  \end{minipage}
  \begin{minipage}[t]{.38\linewidth}\vspace{0pt}
    \typerule{t-uloc-list}{
      \Sigma(\ell) = (n+1, \ext*{\mathsf{list}}, T, \qbot) \\ \ell \subq \flt
    }{
      \G[\flt] \ts \ell : \ext{\ty[\ell]{\TRefpow{n+1}{(\TList~\ty[\qbot]{T})}}}\ \PURE
    }
  \end{minipage}%
  \begin{minipage}[t]{.4\linewidth}\vspace{0pt}
    \typerule{t-uref-list}{
      \G[\flt]\ts t : \ty[d]{(\TList~\ty[\qbot]{T})}\ \EPS
    }{
      \G[\flt]\ts \turef~t : \ty[\QFresh]{(\TRef~\ty[\QFresh]{(\TList~\ty[\qbot]{T})})}\\
      \EPS\EFFSEQ\JUSTM{d}
    }
  \end{minipage}%
  \begin{minipage}[t]{.22\linewidth}\vspace{0pt}
    \typerule{s-list}{
      \G\ts\ty[\qbot]{S} <: \ty[\qbot]{T} \\ \G\ts\ty[\qbot]{T} <: \ty[\qbot]{S}
    }{
      \G\ts\ext{\TList~\ty[\qbot]{S}} <: \ext{\TList~\ty[\qbot]{T}}
    }
  \end{minipage}
  \begin{minipage}[t]{\linewidth}\vspace{0pt}
    \typerule{t-match}{
      \G[\flt]\ts t_1 : \ty[d_1]{(\TList~\ty[\qbot]{T})}\ \EPS[1] \quad
      \G[\flt]\ts t_2 : \ty[d_2]{T_2}\ \EPS[2] \quad \QFresh \notin d_1 \quad
      \cx[\flt, x, y]{(\G, x : \ty[\qbot]{T}, y : \ty[d_1]{(\TRef~\ty[\QFresh]{(\TList~\ty[\qbot]{T})})})} \ts t_3 : \ty[d_3]{T_2}\ \EPS[3] \\
    }{
      \G[\flt]\ts \tmatchlist{t_1}{t_2}{x~y \to t_3} : (\ty[d_2, d_3]{T_2}\ \EPS[1]\EFFSEQ(\EPS[2]\EFFJOIN\EPS[3])) [d_1/y, \qbot/x]
    }
  \end{minipage}

\judgement{Well-Typed Stores (Lists)}{\BOX{\cx[\varphi]{[\Gamma \mid \Sigma]} \mid k \vdash \sigma}}
\[\begin{array}{r@{\quad}l}
    \cx[\varphi]{[\Gamma \mid \Sigma]} \mid k \vdash \sigma := 
    & \DOM(\sigma) \subseteq \DOM(\Sigma) \land |\sigma| = |\Sigma| \land (\varphi \setminus k \subseteq \DOM(\sigma)) \land (k \subseteq \DOM(\Sigma)) \land \mbox{} \\
    & \forall \ell.\ (\ell \in k \implies \sigma(\ell) = \DEAD) \land \mbox{} \\
    & \phantom{\forall \ell.\ } \bigl(\ell \in \varphi \setminus k \land \Sigma(\ell) = (n, \mathit{flag}, T, q) \land \sigma(\ell) = v \implies \mathit{flag} = \mathsf{ref} \land \cx[\varphi]{[\Gamma \mid \Sigma]} \vdash v : \ty[q]{T}\bigr) \land \mbox{} \\
    & \phantom{\forall \ell.\ } \bigl(\ell \in \varphi \setminus k \land \Sigma(\ell) = (n, \mathit{flag}, T, q) \land \sigma(\ell) = v_L \\
    & \phantom{\forall \ell.\ \bigl(\ell \in \varphi \setminus k \land \Sigma(\ell)} \implies \mathit{flag} = \mathsf{list} \land \forall v' \in v_L.\ \cx[\varphi]{[\Gamma \mid \Sigma]} \vdash v' : \ty[q]{T}\bigr) \\
\end{array}\]

\judgement{Reduction Rules (Lists)}{\BOX{\sigma \mid t \redv \sigma\mid t}}
\semrulelabel{step-match-nil} \semrulelabel{step-match-cons}
\[\begin{array}{r@{\,\mid\,}l@{\;}c@{\;}r@{\,\mid\,}l@{\;}r}
  \sigma & \CX[gray]{E}{\tmatchlist{\tnil}{t_2}{x~y \to t_3}} & \redv & \sigma & \CX[gray]{E}{t_2} & \rulename*{step-match-nil} \\ [1ex]
  \sigma & \CX[gray]{E}{\tmatchlist{(\tcons~v~\ell)}{t_2}{x~y \to t_3}} & \redv & \sigma & \CX[gray]{E}{t_3[v/x, \ell/y]} & \rulename*{step-match-cons} \\ [1ex]

\end{array}\]

\semrulelabel{step-swap-nn} \semrulelabel{step-swap-nc} \semrulelabel{step-swap-cn} \semrulelabel{step-swap-cc}
\[\begin{array}{r@{\,\mid\,}l@{\;}c@{\;}r@{\,\mid\,}l@{\;}r}
  \CX[gray]{S}{\ell : []} & \CX[gray]{E}{\tswap~\ell~\tnil} & \redv & \CX[gray]{S}{\ell : []} & \CX[gray]{E}{\tnil} & \rulename*{step-swap-nn} \\ [1ex]
  \CX[gray]{S}{\ell_1 : []}\CX[gray]{S}{\ell_2 : v_L} & \CX[gray]{E}{\tswap~\ell_1~(\tcons~v_2~\ell_2)} & \redv & \multicolumn{3}{l}{} \\
  \multicolumn{4}{r@{\,\mid\,}}{\CX[gray]{S}{\ell_1 : v_2 :: v_L}\CX[gray]{S}{\ell_2 : \DEAD}} & \CX[gray]{E}{\tnil} & \rulename*{step-swap-nc} \\ [1ex]
  \CX[gray]{S}{\ell_1 : v_1 :: v_L} & \CX[gray]{E}{\tswap~\ell_1~\tnil} & \redv & \multicolumn{3}{l}{} \\
  \multicolumn{4}{r@{\,\mid\,}}{\CX[gray]{S}{\ell_1 : []}\,\ell_2 : v_L} & \CX[gray]{E}{\tcons~v_1~\ell_2},\ \ell_2 \not\in \DOM(\sigma) & \rulename*{step-swap-cn} \\ [1ex]
  \CX[gray]{S}{\ell_1 : v_1 :: v_{L1},\ell_4 : v_{L4}} & \CX[gray]{E}{\tswap~\ell_1~(\tcons~v_3~\ell_4)} & \redv & \multicolumn{3}{l}{} \\
  \multicolumn{4}{r@{\,\mid\,}}{\CX[gray]{S}{\ell_1 : v_3 :: v_{L4},\ell_4 : \DEAD}\,\ell_2 : v_{L1}} & \CX[gray]{E}{\tcons~v_1~\ell_2},\ \ell_2 \not\in \DOM(\sigma) & \rulename*{step-swap-cc} \\ [1ex]
\end{array}\]\vspace{-8pt}

\caption{Typing rules, well-typed stores, and operational semantics for unique linked lists. We
highlight the \ext{\text{list-specific types}} and \ext*{\text{list values}}
that distinguish lists from regular locations. }
\label{fig:maybe-list}
\vspace{-3ex}
\end{figure} 
To formalize this, we extend the store typing to support unique lists, enabling
the tracking of potential list elements. Because the operational semantics for
lists are distinct from regular references, we employ specialized terms and typing 
rules for them. However, the underlying typing mechanism remains largely 
consistent with that of unique references. The well-typed store definition in 
\Cref{fig:maybe-list} formalizes this by establishing a correspondence: the 
flag in a location's store type must match the kind of value (a regular value 
for \textsf{ref} or a list value for \textsf{list}) actually present in the 
store at that location.

We introduce \(\TList~\ty[\qbot]{T}\) to model the list structure.\footnote{Because
the current calculus lacks a general recursive type, we introduce \texttt{List} as an
opaque primitive type for simplicity. However, this is purely a formal convenience; in
principle, lists can be fully expressed via Church encoding using recursive function
types.} For simplicity in our formalization and to focus on the core effect mechanism,
we restrict the list extension to only support untracked payloads (\ie, elements with
the \(\qbot\) qualifier). 
\(\TList~\ty[\qbot]{T}\) represents a list value that is either empty (\(\tnil\)) or
a cons cell. In the store, a location of type \(\TList~\ty[\qbot]{T}\) may contain
either an empty list \(\tnil\) or a cons cell \(\tcons~v~\ell\), where \(\ell\)
is a reference to the tail. The \(\tnil\) constructor produces a
\(\TList~\ty[\qbot]{T}\), while \(\tcons\) produces a \(\TList~\ty[\qbot]{T}\)
given a head element and a tail reference. We overload \(\turef\) to allocate a
list value, producing a \(\TRef~\ty[\QFresh]{(\TList~\ty[\qbot]{T})}\). The
\(\tmatchlist{t_1}{t_2}{x~y \to t_3}\) construct destructs a list value \(t_1\). If
\(t_1\) is \(\tnil\), it evaluates \(t_2\). If \(t_1\) is a cons cell, it binds
the head to \(x\) and the tail reference to \(y\) in \(t_3\).
\section{Case Studies} \label{sec:casestudy} 

We illustrate the expressiveness of the effect discipline with three case studies.
We consider the system extended with natural numbers, booleans, and standard
operators such as integer comparison (\eg, less than) and conditionals
(\ie, \texttt{if-then-else}). These extensions follow standard typing
disciplines, including the merging of effects (via the \EFFJOIN\ operator, see
\Cref{fig:effect-ops}). For example, conditionals safely merge effects without
requiring disjointness: since only one branch executes at runtime, the typing
rule
$$
\frac{
  \G[\flt] \ts t_1 : \ty[\qbot]{\TBool}\ \EPS[1] \quad
  \G[\flt] \ts t_2 : \ty[q]{T}\ \EPS[2] \quad
  \G[\flt] \ts t_3 : \ty[r]{T}\ \EPS[3]
}{
  \G[\flt]\ts \tif{t_1}{t_2}{t_3} : \ty[q,r]{T}\ \EPS[1]\EFFSEQ(\EPS[2]\EFFJOIN\EPS[3])
}
$$
simply unions the branch effects ($\EPS[2] \EFFJOIN \EPS[3]$), permitting one branch
to kill a resource while the other merely uses it.

Each case study highlights how the static effect precisely captures the dynamic
behavior proven safe by \Cref{thm:preservation}. We demonstrate traversal and a
destructive \texttt{merge} on unique linked lists, and at-most-once
usage of a regular continuation---a form of ownership transfer through a
higher-order value that complements the freezable counter of
\Cref{sec:motivation-latent}.

\subsection{Operations on Unique Lists}
\label{sec:case-lists}

\subsubsection{Recursive Sum on Unique Lists}
\label{sec:case-sum}

The recursive function in \Cref{fig:case-sum} computes the sum of integers in a
unique linked list. The call @swap(l, nil)@ temporarily detaches the list stored
under @l@ so the program can inspect the branch. In the non-empty branch, the
function recursively processes the tail and then restores the current node with
@swap(l, cons(v, t))@ before returning, so the outer list reference is used but
not consumed.

\begin{figure}[ht]
  \centering
  \begin{varwidth}{\linewidth}
\begin{lstlisting}
def sum(l: Ref[List[Nat]$^\QFresh$]$\trackvar{\QFresh}$): Int$^\qbot$ $\lstcm{\JUSTU{\texttt{l}}}$ =
  val lst = swap(l, nil)                                   // : List[Nat]$^\lstcm{\QFresh}$ $\lstcm{\JUSTU{\texttt{l}}}$
  match lst with 
    nil -> 0                                               
    cons v t -> val s = v + sum(t); swap(l, cons(v, t)); s // $\lstcm{\JUSTU{\texttt{t}}} \EFFSEQ \UKX{\texttt{l,t}}{\texttt{t}}$
  end
\end{lstlisting}
  \end{varwidth}
  \caption{Recursive sum on unique lists.}
  \label{fig:case-sum}
\end{figure}

\subsubsection{Destructive Merge}
\label{sec:case-merge}

\Cref{fig:case-merge} shows \texttt{merge}, a merge that combines two sorted
unique linked lists destructively by rearranging the nodes. The algorithm relies
on \tswap\ and \tmatchlist\ to destructively merge sub-lists, verifying our
static effect system's capability to track the aliases consumed by this merge.
The merge consumes both input lists: empty lists are freed, survivor lists are
restored before they are moved, and recursive calls consume only the selected
tails or lists. The algorithm rearranges and reuses nodes rather than copying or
deeply deallocating whole lists.

Crucially, this destructive consumption of \texttt{l1} and \texttt{l2} is what
lets the result \texttt{merged} emerge as their sole remaining access path---an
instance of the same \emph{derived} exclusivity from \Cref{sec:intro-uniqueness}:
uniqueness is not a primitive annotation on \texttt{merge}'s signature, but a
consequence of the effect discipline tracking that both inputs are fully killed:

\begin{lstlisting}
  val merged = merge(xs)(ys) // : Ref[List[Nat]$^\lstcm{\QFresh}$]$^\lstcm{\QFresh}$ $\lstcm{\UKX{\texttt{xs,ys}}{\texttt{xs,ys}}}$
\end{lstlisting}

\noindent As shown in \Cref{fig:list-ownership}, \texttt{xs} and \texttt{ys} are
both moved into the call, so no alias to either survives; \texttt{merged} is the
unique remaining access path to the combined list.

\begin{figure}[ht]
  \centering
  \begin{varwidth}{\linewidth}
\begin{lstlisting}
def merge(l1: Ref[List[Nat]$^\QFresh$]$\trackvar{\QFresh}$)(l2: Ref[List[Nat]$^\QFresh$]$\trackvar{\QFresh}$): Ref[List[Nat]$^\QFresh$]$\trackvar{\QFresh}$ $\lstcm{\UKX{\texttt{l1,l2}}{\texttt{l1,l2}}}$ =
  val lst1 = swap(l1, nil)                                       // : List[Nat]$^\lstcm{\QFresh}$ $\lstcm{\JUSTU{\texttt{l1}}}$
  match lst1 with
    nil -> free(l1); move(l2)                                    // $\lstcm{\JUSTK{\texttt{l1}} \EFFSEQ \JUSTM{\texttt{l2}}}$
    cons v1 t1 ->
      val lst2 = swap(l2, nil)                                   // : List[Nat]$^\lstcm{\QFresh}$ $\lstcm{\JUSTU{\texttt{l2}}}$
      match lst2 with
        nil -> free(l2); swap(l1, cons(v1, t1)); move(l1)        // $\lstcm{\JUSTK{\texttt{l2}} \EFFSEQ \JUSTM{\texttt{l1}}}$
        cons v2 t2 ->
          if (v1 <= v2) {
            swap(l2, cons(v2, t2)); val rest = merge(t1)(l2)     // $\lstcm{\JUSTM{\texttt{t1,l2}}}$
            swap(l1, cons(v1, rest)); move(l1)                   // $\lstcm{\JUSTM{\texttt{l1}}}$
          } else {
            swap(l1, cons(v1, t1)); val rest = merge(l1)(t2)     // $\lstcm{\JUSTM{\texttt{l1,t2}}}$
            swap(l2, cons(v2, rest)); move(l2)                   // $\lstcm{\JUSTM{\texttt{l2}}}$
          }
      end
  end
\end{lstlisting}
  \end{varwidth}
  \caption{Destructive \texttt{merge} on two unique lists. The algorithm
  consumes both input lists and rearranges/reuses existing nodes rather than
  copying whole lists, so the result is their sole remaining owner.}
  \label{fig:case-merge}
\end{figure}

\subsection{Use-Once Continuations}
\label{sec:case-useonce}

Continuations are notoriously delicate when combined with affine resources: a
single misuse can duplicate or resurrect a capability. This is another instance
of ownership transfer through a higher-order value---here the value is a
continuation.  A reusable continuation has a \emph{pure} latent effect and may
be invoked any number of times; the wrapper below refines that interface so that
invocation \emph{consumes} it. The wrapper @resumeOnce@ in
\Cref{fig:case-resume-once} takes a reusable continuation and exposes it with a
one-shot interface via function latent effects:

\begin{figure}[ht]
  \centering
  \begin{varwidth}{\linewidth}
\begin{lstlisting}
  // A regular continuation cont that can be invoked multiple times
  val cont = ...             // : (V$^\lstcm{\QFresh}$ =>$^\PURE$ Unit$\lstcm{^\qbot}$)
  // Build a one-shot wrapper resumeOnce around cont.
  def resumeOnce[V](k: (V$^\QFresh$ => Unit$^\qbot$)$^\QFresh$) : (f:(v: V$^\QFresh$ =>$^\JUSTM{\texttt{v,f}}$ Unit$^\qbot$))$^{\texttt{f}}$ $\PURE$ = { 
      (f : (v: V$^\QFresh$) => {
        k(v)       // : Unit$^\lstcm{\qbot}$ $\JUSTM{\texttt{v,f}}$
      })}
\end{lstlisting}
  \end{varwidth}
  \caption{A one-shot wrapper for continuations.}
  \label{fig:case-resume-once}
\end{figure}

The type-and-effect derivation for @resumeOnce@ is a direct application of
\Cref{typing:t-sub} (\Cref{fig:maybe-typing}) and \Cref{typing:s-fun}
(\Cref{fig:maybe-subtyping}) on the function latent effect due to covariance.
We note that the latent effect $\JUSTM{\texttt{v,f}}$ of the returned
continuation kills both the returned continuation itself and the argument passed
to it. As a result, calling @f@ more than once results in a type error:

\needspace{8\baselineskip}
\begin{lstlisting}
  // Use-site: exactly-once behavior
  val v = ...                              // : V$^\lstcm{\texttt{v}}$
  val f = resumeOnce(cont)                 // : (f: (v: V$^\lstcm{\QFresh}$) =>$^\lstcm{\JUSTM{\texttt{v,f}}}$ Unit$^\lstcm{\qbot}$)$^\lstcm{\texttt{f}}$ $\PURE$
  val y = f(v)                             // : Unit$^\lstcm{\qbot}$ $\JUSTM{\texttt{v,f}}$   // consumes f and k
  val z = f(v)                             // $\lsterrcm{\texttt{ill-typed: f and v were killed by the previous call}}$
\end{lstlisting}

Alternatively, we could make @resumeOnce@ more aggressive by letting it kill the
original continuation, so that the continuation can only be accessed via the
returned one-shot wrapper:

\begin{lstlisting}
  val g = resumeOnceAndMove(cont)          // : (f: (v: V$^\lstcm{\QFresh}$) =>$^\lstcm{\JUSTM{\texttt{v,f}}}$ Unit$^\lstcm{\qbot}$)$^\lstcm{\texttt{f}}$ $\JUSTK{\texttt{cont}}$
  val h = cont(v)                          // $\lsterrcm{\texttt{ill-typed: cont is moved into g}}$
\end{lstlisting}

This refinement is stable under reduction: the operational semantics never
deallocate the thunk representing @f@, and it is invoked just as a normal
continuation. However, the static type-and-effect system will prevent it from
being called more than once at runtime, thus mirroring the at-most-once
guarantee that the static type system enforces.

Although we do not alter the runtime behavior in this case, a compiler
implementation can leverage the static type-and-effect information to improve
runtime
performance~\Citep{Proc.42ndACMSIGPLANInt.Conf.Program.Lang.Des.Implement./SivaramakrishnanSLTS21a},
knowing that the continuation can be invoked at most once. In effect,
\texttt{resumeOnce} performs ownership transfer on a higher-order value: the right
to invoke the continuation is moved into the one-shot wrapper and consumed on use.

\section{Limitations and Future Work}
\label{sec:limitation}

\paragraph{Physical In-Place Moves and Swaps}
Our operational semantics implements both \tmove\ and \tswap\ by allocating a fresh
destination and deallocating the source. This is a sound model of ownership
transfer on individual references, but it is not intended to faithfully model a
true in-place move or swap at the physical memory level.
In-place swaps would break the telescoping invariant of the store typing and
require a virtual layer that can directly observe alias reshuffling.

Despite the lack of in-place moves in the operational semantics, our
proofs establish the key desired safety guarantees, namely that locations 
cannot be dereferenced \emph{after} being moved. 
A completely faithful model of in-place moves could be achieved by
adding a \emph{virtual memory} layer in the proofs to model re-allocation
as remapping of a physical location under a new virtual address, thus
maintaining the telescoping invariant on virtual addresses. 
Virtual addresses would remain stable while their targets in the
concrete heap are permuted, allowing swaps to remap aliases 
without copying data and avoiding spurious allocations.

\paragraph{Deep Copy on Non-References}

This restriction concerns moving a closure \emph{value}, not transferring
ownership \emph{through} a closure: the freezable counter
(\Cref{sec:motivation-latent}) and use-once continuations proceed through
latent \textit{use}/\textit{kill} effects and are fully supported. We restrict \tmove, \turef, and \tswap\ to references and lists
because moving a closure itself would have to move or disable every captured
reference transitively, and the store typing for unique references would have
to speak of a \emph{set of locations} that are simultaneously unique from the
rest of the store.

Beyond such bookkeeping, higher-order moves entail a genuine redesign, because
the moved function's \emph{signature} can also mention the moved state---for
instance, when a function returns a captured resource:

\begin{lstlisting}
    val c = new Ref(0)  // : Ref[Int]$^\lstcm{\trackvar{c}}$
    val f = fun () => c // : (Unit => Ref[Int]$\lstcm{\trackvar{c}}$)$\lstcm{\trackvar{c}}$
    val g = move(f)     // : (Unit => Ref[Int]$\lstcm{\trackvar{c}}$)$^\lstcm{\QFresh}$
        // $\lsterrcm{\text{Error: signature still names c}}$
\end{lstlisting}

A parametric \tmove\ would return \texttt{f} at a fresh outer qualifier
\(\QFresh\)---a set of \emph{new} locations---while the unchanged signature
still names the old location \texttt{c}, so the moved value is no longer
typeable and preservation fails. Restoring typeability requires structural
changes: rewriting signatures under moves, so that types change across
reduction; weakening the preservation invariant to account for the discrepancy
between outer and interior qualifiers; or reinterpreting function qualifiers so
that interior mentions resolve through the outer one. Each option
\textit{alters} the core system rather than extending it, so we leave
higher-order moves to future work.

This challenge is a direct consequence of the fine-grained reachability tracking
our calculus provides. Systems that rely on strict topological isolation or
coarse-grained regions bypass this issue simply because a moved value's
signature usually cannot mention the moved location.  In contrast, our system
tracks capturing and permits overlapping with the type signature, so that same
precision also forces the signature to keep naming \texttt{c} after the move.
\section{Related Work}
\label{sec:related}

\paragraph{\textbf{Reachability Types}}

Reachability types (RT) \Citep{DBLP:journals/pacmpl/BaoWBJHR21} qualify expressions
by the variables or locations they may reach, thereby tracking aliasing and
separation in higher-order programs with mutable references.
\Citet{DBLP:journals/pacmpl/WeiBJBR24} extends their work to support
polymorphism and higher-order references.
\Citet{Proc.ACMProgram.Lang./DengSSY25} introduces cyclic reference types,
shallow one-step reference tracking, and dual-component references for
controlled escaping. Their shallow model decouples an outer reference from its
transitively reachable referent, improving precision for nested structures and
fine-grained parallelism. Our work builds on this alias-tracking granularity and
adds a flow-sensitive effect layer for resource destruction and ownership
transfer on individual references.

\Citet{he_when_2025} uses reachability to provide lifetime control for
higher-order programs: arenas permit arbitrary in-arena sharing with statically
scoped reclamation, while individually managed resources remain possible; the
system deliberately avoids flow-sensitive reasoning. In contrast, our system
adds a flow-sensitive destructive effect layer to RT. These two approaches are
complementary rather than mutually exclusive: arenas provide a coarse-grained
allocation and reclamation layer,
while use and kill effects provide fine-grained control for references that must
be reclaimed or transferred before their enclosing arena dies. We intentionally
keep the core system simple and leave the formal combination with arenas to
future work.

Additionally, \Citet{bao_modeling_2025} develop logical-relations foundations
for RT, including termination in the presence of higher-order store. Graph IR
\citep{DBLP:journals/pacmpl/BracevacWJAJBR23} leverages RT to optimize impure
higher-order programs by tracking fine-grained dependencies with an effect
system; its effects are used for dependency analysis rather than destructive
resource operations or ownership transfer. \Citet{jia_escape_2024} address
self-references by proposing an enhanced notion of subtyping and a sound,
decidable bidirectional typing algorithm for RT; this is complementary to our
flow-sensitive reasoning of use and deallocation.

\paragraph{\textbf{Ownership Transfer in Higher-Order Languages}}

Ownership transfer through higher-order functions is not, by itself, new. Linear
and affine systems---such as Linear
Haskell~\citep{DBLP:journals/pacmpl/BernardyBNJS18}, the linearity/uniqueness
account of \citet{ProgrammingLanguagesandSystems/MarshallM22}, and
Alms~\citep{SIGPLANNot./Tov11}---pass and consume resources as first-class
values. Similarly, the permission-based language Mezzo threads exclusive
permissions through closures~\citep{ACMTrans.Program.Lang.Syst./BalabonskiF16},
while capability and \texttt{iso} systems like Pony~\citep{clebsch_deny_2015},
Verona's Reggio~\citep{Proc.ACMProgram.Lang./ArvidssonESSJM23}, and Fearless
Concurrency~\citep{DBLP:conf/pldi/MilanoTM22} transfer isolated object graphs
between contexts. What distinguishes these systems is \emph{how} they secure the
transfer: they either forbid shared mutable aliasing entirely (via linear/affine
consumption or \texttt{iso} isolation) or impose heavy surface annotation
burdens. Consequently, ownership transfer is forced at \emph{capture} time: a
closure capturing a linear value must itself become linear or one-shot (e.g.,
Linear Haskell's multiplicities, Mezzo's permission threading), and Rust
moves a consumed value into the closure's struct at creation
(\Cref{fig:intro-capture-vs-call}). 

Our system addresses this gap by decoupling capture from consumption. Because
closures are never owners, transfer is deferred entirely to \emph{application}.
Rather than relying on primitive attributes like the \texttt{iso} keywords found
in reference-capability systems, we \textit{derive} uniqueness as a mode of use
via flow-sensitive \textit{use}/\textit{kill} effects carried latently on
function types. Furthermore, in contrast to Rust's rigid restrictions, we permit
shared mutable aliasing from the outset, recovering safety through flow-sensitive
revocation. This synthesis of shared aliasing, non-lexical destruction, and
derived uniqueness effortlessly composes with higher-order, impure functions.
Deferred consumption is exactly what permits a \texttt{freeze} closure to retire
a shared counter (\Cref{sec:motivation-latent}) while guaranteeing every
remaining alias stays statically safe. Ultimately, we argue this is the correct
foundation for ownership transfer under higher-order control flow.

\paragraph{\textbf{Region-Based Systems}}

Classic region calculi provide static memory management with explicit
effects~\citep{lippmeier_mechanized_2013}, while region inference historically
pursued early deallocation via whole-program
constraints~\citep{DBLP:conf/popl/TofteT94,DBLP:conf/pldi/AikenFL95,InformationandComputation/Tofte97}.
However, while these foundational systems may track lifetimes via effects, they
rely on the bulk reclamation of entire regions rather than the flow-sensitive
revocation of individual aliases. Among modern approaches, Spegion is closest
in spirit: it uses effects to support non-lexical, safe manual memory management
without substructural types, justifying the deallocation of aliased
regions~\citep{hughes_spegion_2025}. Despite these similarities, Spegion focuses
on C-like buffer reasoning. Our system instead tracks reachability at the
granularity of individual locations, enabling fine-grained ownership transfer,
unique references, and destructive list idioms that Spegion does not model.

Other advanced capability systems rely on rigid region topologies rather than
individual aliases: Verona's Reggio organizes the heap into a forest of
isolated regions with explicit thread windows, mediating transfer and
reclamation entirely through these topological
boundaries~\citep{Proc.ACMProgram.Lang./ArvidssonESSJM23}, while Fearless
Concurrency safely transfers complex object graphs using reservations, tempered
domination, and dominating \texttt{iso} fields, occasionally falling back on
dynamic \texttt{if disconnected} checks~\citep{DBLP:conf/pldi/MilanoTM22}. Both
mandate a prearranged region topology and operate in strictly first-order
contexts, whereas our system assumes none: it supports non-lexical,
programmer-controlled reclamation and ownership transfer directly over
reachability-qualified aliases within a fully first-class, higher-order
functional setting.

\paragraph{\textbf{Separation and Alias Control}}

Separation Logic enables local heap reasoning via separating
conjunction~\citep{DBLP:conf/lics/Reynolds02}. Reachability types (RT) adapt
this principle natively, enforcing separation via disjoint function and argument
qualifiers~\citep{DBLP:journals/pacmpl/BaoWBJHR21,DBLP:journals/pacmpl/WeiBJBR24}.
While concurrent systems extend Capture Calculus
(CC)~\citep{xu_degrees_2023,Proc.ACMProgram.Lang./XuA24,DBLP:journals/toplas/BoruchGruszeckiOLLB23}
or merge separation logic with effect
handlers~\citep{Proc.ACMProgram.Lang./DeVilhena21}, we achieve effect
safety directly. Augmenting RT with destructive effects and move semantics
guarantees safe ownership transfer without region windows or burdensome 
user-supplied separation proofs.

Other approaches manage state using complex type decorations: coeffect systems
track reduction-induced sharing for capsule
invariants~\citep{Proc.ACMProgram.Lang./BianchiniFPE22,TheoreticalComputerScience/BianchiniFP23};
DOT formalizes path-dependent types for mutability and
purity~\citep{DBLP:conf/oopsla/RompfA16,LIPIcsVol.166ECOOP2020/Dort20,38thEur.Conf.Object-OrientedProgram.ECOOP2024/DortYO24};
and qualified types drive inference for immutability and
polymorphism~\citep{ACMTrans.Program.Lang.Syst./FosterRJ06,SIGPLANNot./HuangAW12,Proc.ACMProgram.Lang./Lee23}.
Rather than relying on static capabilities or qualifiers, our RT extension 
explicitly exposes aliasing. Controlling consumption via flow-sensitive 
\textit{use}, \textit{kill}, and \textit{move} effects shifts the analytic 
focus from \emph{what} attributes decorate a resource to precisely \emph{when} 
it may be safely accessed or destroyed.

\paragraph{\textbf{Object and Reference Capabilities}}

Pony combines actors with deny-style reference capabilities for static
data-race freedom~\citep{clebsch_deny_2015}. Contextual modal type
theory~\citep{DBLP:journals/tocl/NanevskiPP08} grounded scoped and capturing
capabilities for System $F_{<:}$, later extended with capture sets,
subcapturing, and boxing in
Scala~\citep{odersky_scoped_2022,DBLP:journals/toplas/BoruchGruszeckiOLLB23},
while other capability frameworks enforce at-most-once consumption alongside
flexible borrowing~\citep{ECOOP2010-Object-OrientedProgram./Haller10}. These
systems control authority, aliasing, or consumption primarily through rigid
capability restrictions; in contrast, our system tracks \textit{use},
\textit{kill}, and \textit{move} via flow-sensitive effects layered over RT's
reachability qualifiers, without static capability constraints.

Substructural systems---such as Linear
Haskell~\citep{DBLP:journals/pacmpl/BernardyBNJS18}, the linearity/uniqueness
synthesis of \citet{ProgrammingLanguagesandSystems/MarshallM22}, and fully
in-place functional execution via linear
calculi~\citep{Proc.ACMProgram.Lang./LorenzenD23}---control consumption through
substructural or uniqueness disciplines. These approaches inherently restrict
aliasing to enforce safety, whereas our effect discipline leaves aliasing
explicitly visible and simply rejects any subsequent uses after a resource
undergoes destruction or transfer.

\paragraph{\textbf{Memory Management}}

Cyclone secures manual memory management in a C-like setting via lexical and
dynamic regions, alongside unique pointers for flow-sensitively checked
individual
deallocation~\citep{Proc.ACMSIGPLAN2002Conf.Program.Lang.Des.Implement./GrossmanGTMY02,DBLP:conf/iwmm/HicksMGJ04}.
Crucially, its object-level \texttt{free} requires uniqueness or region
boundaries rather than revoking shared aliases. Pony's ORCA collector leverages
type-level capability restrictions for independent actor garbage collection
rather than flow-sensitive revocation~\citep{clebsch2015ownership}. Rust moves
by default and supports explicit early destruction (\texttt{std::mem::drop})
alongside replacement idioms for partial moves (\texttt{replace},
\texttt{take}); however, these manual deallocation operations demand an
exclusively mutable path and cannot conflict with active aliases or
borrows~\citep{klabnik2019rust,DBLP:journals/pacmpl/0002JKD18}.

These systems enforce a strict dichotomy: developers must choose between shared
mutability and deterministic, manual deallocation, but they cannot have both.
Our system resolves this tension by tracking shared mutable aliases in
higher-order functions via RT and effects. We enable flow-sensitive destruction
and ownership transfer on individual references even while they remain shared.
Deep, whole-graph ownership transfer remains orthogonal and recoverable via an
arena layer~\citep{he_when_2025} or dual-component references that expose inner
reachability~\citep{Proc.ACMProgram.Lang./DengSSY25}. %
\section{Conclusion}%
\label{sec:conclusion}

We presented a flow-sensitive type-and-effect system for reachability types that
makes ownership transfer sound in the presence of higher-order control flow while
retaining shared mutable aliasing. Its two effects, \textit{use} and \textit{kill},
yield no-use-after-free/kill guarantees, ownership transfer on individual
references, and uniqueness as a derived property, with destruction and consumption
expressible \emph{through} closures. We prove effect soundness (progress,
preservation, multi-step and parallel reduction) in Rocq, with case studies
including a destructive merge on unique lists, a freezable counter, and
use-once continuations.

\section*{Data Availability Statement} An anonymized archive of the development
accompanies this submission and will be released alongside the camera-ready
artifact.

\bibliography{references}

\end{document}